# When to Update Systematic Literature Reviews in Software Engineering

Emilia Mendes[1], Claes Wohlin[2], Katia Felizardo[3], Marcos Kalinowski[4]

[1]Department of Computer Science Blekinge Institute of Technology Karlskrona, Sweden emilia.mendes@bth.se

[2] Department of Software Engineering Blekinge Institute of Technology Karlskrona, Sweden claes.wohlin@bth.se

[3] Department of Computing Federal Technological University of Paraná Cornélio Procópio, Brazil katiascannavino@utfpr.edu.br

[4] Department of Informatics Pontifical Catholic University of Rio de Janeiro (PUC-Rio) Rio de Janeiro, Brazil kalinowski@inf.puc-rio.br

**Abstract.** [*Context*] Systematic Literature Reviews (SLRs) have been adopted by the Software Engineering (SE) community for approximately 15 years to provide meaningful summaries of evidence on several topics. Many of these SLRs are now potentially outdated, and there are no systematic proposals on when to update SLRs in SE. [*Objective*] The goal of this paper is to provide recommendations on when to update SLRs in SE. [*Method*] We evaluated, using a three-step approach, a third-party decision framework (3PDF) employed in other fields, to decide whether SLRs need updating. First, we conducted a literature review of SLR updates in SE and contacted the authors to obtain their feedback relating to the usefulness of the 3PDF within the context of SLR updates in SE. Second, we used these authors' feedback to see whether the framework needed any adaptation; none was suggested. Third, we applied the 3PDF to the SLR updates identified in our literature review. [*Results*] The 3PDF showed that 14 of the 20 SLRs did not need updating. This supports the use of a decision support mechanism (such as the 3PDF) to help the SE community decide when to update SLRs. [*Conclusions*] We put forward that the 3PDF should be adopted by the SE community to keep relevant evidence up to date and to avoid wasting effort with unnecessary updates.

# 1 Introduction

In their call to arms paper, Kitchenham et al. [13] argued for an Evidence-Based paradigm in Software Engineering (EBSE), to be particularly employed by *"researchers interested in empirical software engineering and practitioners faced with decisions about the adoption of new software engineering technologies."* Furthermore, they state that EBSE's goal is: *"to provide the means by which current best evidence from research can be integrated with practical experience and human values in the decision-making process regarding the development and maintenance of software"*, and also encourage the use of systematic literature reviews (SLRs) to obtain such current best evidence.

By mid-2008, 53 SLRs have been published in Software Engineering (SE) (20 found via a manual search of ten journals and four conferences [10], and the remaining 33 via searches in six databases [11]). Another 67 [28] SLRs were identified between January and December 2009; and finally an additional 316 SLRs between January 2010 and May 2016 [4], thus leading to a total of 436+ SLRs published in SE, within the period from January 2004 to May 2016. Furthermore, as detailed in Section 3, our database searches identified 20 updated SLRs. It provides evidence that many SLRs in SE are potentially outdated; thus affecting our current aggregated understanding of the state-of-the-art in those SLRs' research topics. Note that if we also consider potentially outdated primary studies, e.g. studies that refer to a particular technology that is not used any longer, the situation is even more challenging.

When new evidence is added as part of updating SLRs, different findings and conclusions from those reported initially may be identified. Therefore, updating SLRs may contribute to different purposes. For example, (a) providing a continuous update of the state-of-the-art on a research topic; and (b) identifying how that particular research topic is evolving (e.g. are the overall conclusions from previous SLRs stable throughout different SLR updates, or are they changing? If the latter, are these changes significant enough to make previous SLRs' findings worth updating?).

How to keep SLRs current, more specifically concerning how and when to update SLRs, has also been a topic of discussion in other fields, such as medicine (e.g. [16][21][23][59]) and environmental sciences (e.g. [34]). Although Cochrane has advocated in the past that SLRs in Medicine and other Healthcare specialities should be updated every two years [39], the guide, published in 2008 [64], argued for a more detailed decision-making process, and provided a decision tool to guide such actions. In 2014, Cochrane organised a two-day workshop, attended by *"an international panel of authors, editors, clinicians, statisticians, information specialists, other methodologists, and guideline developers"*, aimed to identify *"a common approach to updating systematic reviews, drawing on existing strategies, research, and experience of people working in this area"*[1]. The results from this workshop were published in 2016 [58], where they also included a decision framework to guide the update (or not) of SLRs in a wide range of fields, not only Medicine.

Within the context of SE, SLRs also need updating; however, we have only found a short paper discussing this issue, and using the term SLR maintenance, rather than SLR update [75]. They surveyed 28 authors of SLRs (not SLR updates) to identify the relevance to keep SLRs up-to-date and report that approximately 2/3 of those 28 respondents would like to keep their SLRs up-to-date. These results support the need for a decision framework or detailed guidelines to assess whether an SLR ought to be updated. Software Engineering can benefit from learning from other disciplines that are already doing this systematically. Therefore, in a similar way as Kitchenham [7] and Kitchenham and Charters [12], who used knowledge from more mature disciplines (in this regard, the knowledge relates to existing guidelines for conducting SLRs used by medical and social sciences researchers) to propose guidelines for conducting SLRs in SE, we have investigated existing decision frameworks and detailed guidelines in other more mature disciplines with respect to SLRs ([16][21][23][33][59][65]). We put forward that the decision framework proposed in [58] – 3PDF henceforth, can be used as a basis to guide the decision of when to update SLRs in SE. Section 2 introduces some existing decision frameworks in other disciplines, and details the 3PDF Framework, which is the one applied to assess SLR updates in SE (see Section 5). This is one of the main contributions of this paper.

Furthermore, we also argue that the definition of an SLR update, as suggested by Garner et al. [58], is broad enough to be adopted within the context of SE. Note that the definition resulted from discussions led by a panel of experts who carried out a variety of SLRs in a wide range of fields, not only restricted to Medicine and other Healthcare-related fields[1].

This panel's definition is as follows:

> *"an update of a systematic review [is defined] as a new edition of a published systematic review with changes that can include new data, new methods, or new analyses to the previous edition … An update asks a similar question with regard to the participants, intervention, comparisons, and outcomes (PICO) and has similar objectives; thus, it has similar inclusion criteria. These inclusion criteria can be modified in the light of developments within the topic area with new interventions, new standards, and new approaches. Updates will include a new search for potentially relevant studies and incorporate any eligible studies or data; and adjust the findings and conclusions as appropriate."*

Within the context of this definition, new data represents new primary studies found via new searches – manual and/or in databases; new methods are, for example, to employ more detailed quality checklists to assess evidence, or to provide additional ways to aggregate evidence (e.g. thematic analysis), or to employ different ways to identify primary studies (e.g. database search vs snowballing[2]); finally, new analyses may include re-doing the data analysis based on, for example, more recent meta-analysis and research synthesis methods.

The specific goals and main contributions of this paper, which relate to when to update SLRs in SE, are fourfold: first, to identify existing policies, guidelines, recommendations relating to when to update SLRs to check whether they can be used within the context of SE; second, to carry out a literature review to identify SLR updates in SE; third, to carry out a survey aimed at the authors of these SE SLR updates; the survey's goal is to obtain these authors' feedback about using the policy/guideline/recommendation identified as part of our first research goal, as possible decision mechanism to support the decision about when/if to update an SLR in SE; and finally fourth, to apply the decision mechanism outcome of our third research goal to assess already published SE SLR updates (obtained as part of our second research goal). Based on our

---

[2] Note that within the context of this paper the term 'database search' represents a string-based automated search carried out on a database (e.g. SCOPUS) and/or Digital library (e.g. IEEEXplore, ACM). 'Backward snowballing' represents both a manual search over a paper's reference list and a string-based automated search on a database or digital library. Finally, 'forward snowballing' represents an automated search based on citation analysis.

results, we provide recommendations on when to update SLRs in SE. These four goals are translated into four overall research questions (ORQ), addressed respectively in Sections 2 to 5:

ORQ1: What are (if applicable) the policies, guidelines, recommendations concerning when to update SLRs?

ORQ2: How have SLR updates been conducted in software engineering?

ORQ3: What is the perceived usefulness of the decision mechanism identified in ORQ1 to the authors of the SE SLR updates identified via ORQ2?

ORQ4: If using the decision mechanism (ORQ3), which SLR updates in SE (ORQ2) would be assessed as suitable for an update?

The remainder of this paper is organised as follows. In Section 2, we present some previous work in other fields relating to when to update SLRs and further describe the 3PDF framework. Section 3 reports a literature review of SLR updates in SE. In Section 4, we detail a survey employed to obtain feedback from the authors of the SLR updates presented in Section 3. Section 5 presents the use of the 3PDF to the SLR updates from Section 3. Section 6 contains a discussion on threats to validity, followed by an argumentation on research implications in Section 7. Finally, our conclusions and comments on future work are given in Section 8.

## 2  Policies, Guidelines or Recommendations on When to Update SLRs

### 2.1  Previous proposals

This sub-section introduces seven articles that were identified via searches carried out by the first author on Google Scholar and Scopus, using the following strings: "*When+and+how+to+update+systematic+reviews*", and "*update systematic reviews*". A total of 298 references were screened in Google Scholar and 35 in Scopus; seven and five references were selected, respectively. Note that after duplicates were removed, the final number of papers was seven. Also note that we found two different references to the same paper, which was published as an article [22] and also as a document in the Cochrane Database of Systematic Reviews 2008 [23]. This search was not intended to be an SLR, and therefore we used only Google Scholar and Scopus. Except for one study, which was in the field of environmental management [33], all remaining studies were in the medical field.

Studies were arranged into three categories: i) systematic review about when and how to update SLRs; ii) techniques to identify new evidence related to previously published SLRs; and iii) decision mechanisms (factors or decision tree) to decide whether an SLR needs updating.

## Systematic review about when and how to update SLRs

We found one article in this category, which had been published in two different places [22][23]. Its authors looked for previous work (fully refereed papers and also grey literature) describing strategies and methods addressing when and/or how to update SLRs. They searched some databases - MEDLINE (from 1966 to 2005), PsycINFO, and also the Cochrane Methodology Register (Issue1, 2006), and manually searched the 2005 Cochrane Colloquium proceedings. They screened 2548 records and included 15 studies that characterized four strategies, one technique and two statistical methods describing when or how to update SLRs. We refer interested readers to Tables 1, 2, and 3 in [22]. Their assessment shows several significant limitations throughout all proposals, and their overall conclusion is that "The identified strategies are not pragmatic and have not been empirically tested. More concerted research efforts are needed to illuminate further knowledge gaps in the field of updating SRs and to ascertain the potential benefits of developing internationally harmonized, efficient, yet valid, ways of updating SRs.".

## Techniques to identify new evidence related to previously published SLRs

Three studies assumed that the issue about when to update SLRs was solely driven by the existence of more recent evidence relating to an SLR [48][43][59]. Two studies were only based on the results from manual database searches [48][43]; the third study used a combination of manual searches and statistical techniques that assessed whether new evidence would change the conclusions from meta-analyses based upon evidence included in the perceived outdated SLRs [59]. In this latter case, any SLRs candidate for update should also include meta-analyses.

## Decision mechanisms (factors or decision tree) to decide whether an SLR needs updating

The first study to present a decision mechanism to decide whether an SLR was eligible for updating was carried out by Takwoingi et al. [77]. They proposed a decision tool, which in fact can be seen as a precursor to Garner et al.'s decision tool [58] (presented in detail in sub-section 2.2). This decision tool supports the judgement as to whether an SLR is qualified for updating via three consecutive steps (detailed in Figure 2 in their paper):

"Step 1: Is the clinical question already answered by the available evidence or is the clinical question deemed no longer relevant?

Step 2: Are there any new factors relevant to the existing review? (e.g. additional information from existing included studies, previously missing data)

Step 3: Are there new studies?"

The tool assumes that all SLRs should both relate to clinical trials and include a meta-analysis. Furthermore, SLRs are judged as not in need for updating when there are new factors to consider but when applying the new factor(s) there is no new evidence and/or new conclusions. Overall, this decision tool is much more limiting when compared to Garner et al.'s decision tool, which, amongst other things, was proposed for use within and outside the medical field. Note that the paper's two first authors were also part of the team that participated in Garner et al.'s study and workshop, which is also detailed in Sub-section 2.2.

The second study, by Bayliss et al. [33], proposes factors that are believed to affect the decision as to whether an SLR update is needed. The set of factors (listed below) is not arranged as part of a decision tool but simply listed. Furthermore, the proposal is much simpler than the ones by Takwoingi et al. [77] and Garner et al. [58] and focuses on whether new evidence on the SLR's topic has been published. The factors are as follows:

a. The SLR's topic is still relevant to policy and practice, and some preliminary searches suggest that there are new studies suitable for inclusion.

b. The SLR's topic is relatively new, and the original SLR included limited data. An update should be done if it is acknowledged that such an update would provide valuable additional information.

c. "Where large volumes of information have been published over a short timescale, for example for topical issues (e.g. pollinators) or broad areas (e.g. climate change), especially where evidence is contentious or where trends in research methods, study subjects, research groups or dogma have demonstrated a shift over time—new evidence may be more likely to contradict old."

d. If large influential studies were published and may affect the original SLR's conclusions.

The third study, by Garner et al. [58], details a decision tool, arranged as a decision tree, which considers different aspects relating to the suitability for an SLR to be updated (not only the existence of new evidence). The proposal, to be detailed in the next sub-section, is more comprehensive than all the other suggestions abovementioned, and also aims to support decisions in a wider range of fields outside the medical field.

## 2.2 Decision framework proposed by Garner et al.

Garner et al.'s [58] decision framework was selected within the context of this research for several reasons. One of these reasons has already been mentioned in Sub-section 2.1 and relates to including decision points that go beyond the existence of newly published evidence on the SLR's topic. The framework, which is presented in Figure 1, also supports the updating of SLRs based upon other considerations such as the use of new relevant methods. Another reason for choosing this framework is that it is the only study amongst those introduced in Sub-section 2.1 that targets an audience both within and outside the medical field. Therefore, we believe that it is generic enough so to also be applicable within the SE context.

Furthermore, Garner et al.'s decision framework resulted from a collaboration between different stakeholders aiming to put forward as decision support to update SLRs in different areas, by drawing on existing *"strategies, research, and experience of people working in this area"*[58]. Therefore, it is an improvement over the tools that had been previously suggested by Cochrane and others. Note that Section 4 describes a survey in which feedback about the use of this framework within SE is discussed. Moreover, we also evaluate this framework by applying it to twenty SLR updates in SE.

The decision framework includes three steps (See Figure 1), to be applied sequentially, as follows.

**Step 1** focuses on assessing how current/actual the SLR is by looking at its topic's relevance for research and practice. This includes whether the published SLR has had an impact on research and/or practice (using metrics such as citations via sites such as Google Scholar), and finally whether the SLR was carried out properly and using a sound methodology (e.g. clear inclusion/exclusion criteria and quality assessment). As shown in Figure 1, if the answer to at least one of Step 1's questions is 'NO', there is no arguable need for updating this SLR. Note that in the particular case where questions a. and b. are 'YES' and question c. 'NO', an update of the original SLR is not warranted; however the original SLR should in fact be redone given that the method originally used is not viewed as valid. While being redone, it can also widen its search so to include more up to date primary studies. The above mentioned framework does not target this situation, given that it is not an update, but a new SLR. Possible answers are 'YES' and 'NO'.

**Step 2** focuses on identifying if there are any new methods proposed and/or new studies published after the SLR's publication. Examples of new methods can be new ways to search for evidence and to do the synthesis of evidence, the inclusion of other types of study designs (e.g. qualitative research), new statistical techniques or improvements to existing techniques used for meta-analysis. Step 2.b relates specifically to whether new additional studies were found. In this step, as long as at least one question is affirmative, one can proceed to the next step – Step 3.  Possible answers are 'YES' and 'NO'.

Finally, **Step 3** targets at assessing whether the adoption of new methods and/or new studies may affect the conclusions when compared to the conclusions from the original SLR, and/or the original SLR's credibility. Note that concerning 3.b, there is a range of different aspects to consider. For example, conclusions may change due to adding new primary studies to the original SLR. These would be identified by looking at the results from database searches/snowballing, based on reading titles and abstracts. Note that such a task will most likely be positively impacted whenever prospective studies use structured abstracts. Other examples would be changes made to the inclusion criteria or quality checklist which, when reapplied to the studies in the original SLR, lead to changes in the conclusions. The same applies here as in Step 2, i.e., as long as at least one question is affirmative, it is recommended to proceed to the next step. Possible answers are 'YES/MAYBE' and 'NO'.

At the end of Step 3, it is clear that an SLR needs to be updated. Suggestions on how to update such SLRs are provided in Section 5, and details on how to carry out searches are given in Mendes et al. [26].

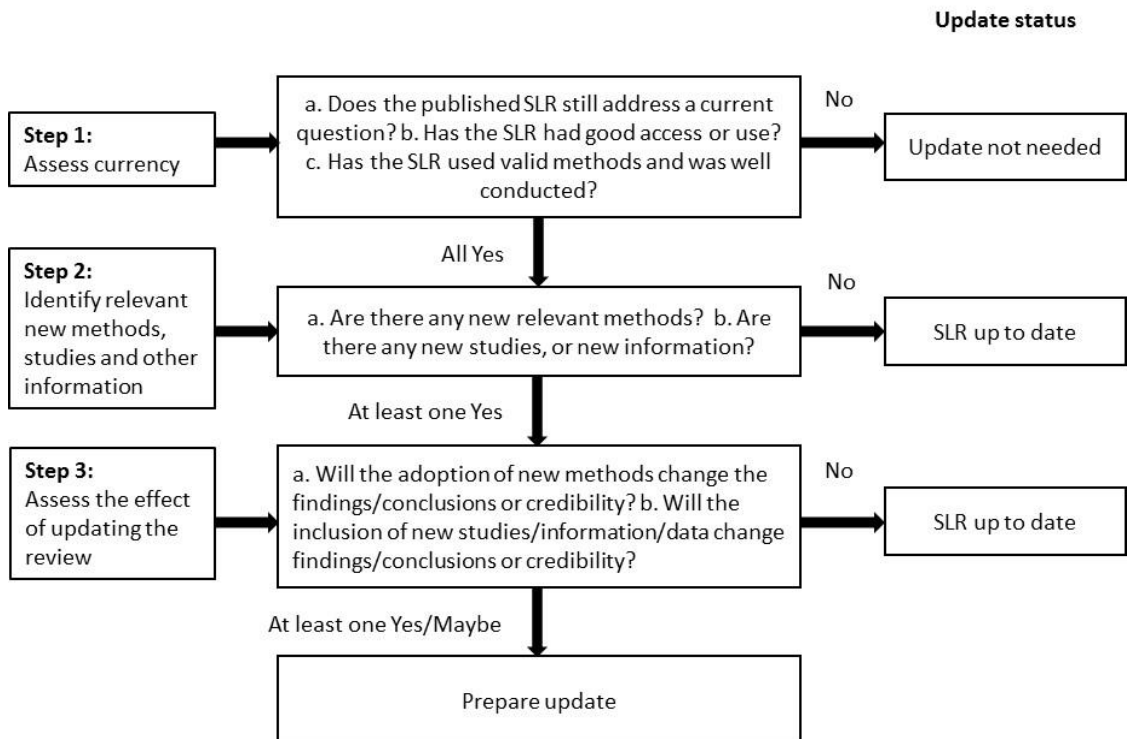

**Figure 1: Decision framework to assess SLRs for updating [58].**

# 3 Searching for existing SLR updates in SE

This section details a Systematic Literature Review (SLR) carried out to identify SLR updates in Software Engineering (SE). The reasons for carrying out this SLR were three-fold: first, to identify a sample of SLRs updates in SE. We use the word sample because no SLR that has as one of its inclusion criterion "the study is written in English" can claim full coverage of studies. Second, to identify SE SLR update authors. These authors are invited later to participate in a survey aimed at obtaining their feedback in relation to: i) the adoption of the 3PDF framework within the context of SE SLR updates; and ii) the framework itself. This survey is detailed in Section 4; and third, to apply the 3PDF framework (and any recommended changes resulting from the survey investigation) to each of the SE SLR updates presented in this section (this is detailed in Section 5).

**Research Questions**

There are seven Research Questions (RQs), in addition to the four overall research questions (ORQ), we are interested in, which are directly related to the three Steps used in the 3PDF framework when deciding whether an SLR needs to be updated. We have matched the RQs with the framework steps because these drove both the data extraction phase and the results presented in this section.

The RQs are as follows:

- **RQ1** (relates to Step 1.a. in the 3PDF): Does the published SLR still address a current question?
- **RQ2** (relates to Step 1.b. in the 3PDF): Has the SLR had good access or use?
- **RQ3** (relates to Step 1.c. in the 3PDF): Has the SLR used valid methods and was well-conducted?
- **RQ4** (relates to Step 2.a. in the 3PDF): Are there any new relevant methods?
- **RQ5** (relates to Step 2.b. in the 3PDF): Are there any new studies, or new information?
- **RQ6** (relates to Step 3.a. in the 3PDF): Will the adoption of new methods change the findings, conclusions or credibility?
- **RQ7** (relates to Step 3.b. in the 3PDF): Will the inclusion of new studies/information/data change findings, conclusions or credibility?

**Inclusion and Exclusion Criteria**

We used as inclusion criteria (IC):

- **IC1**: The study is an updated SLR; and
- **IC2**: Both the original secondary study and its update are available, and provide the data that is needed for comparison, e.g., the search string used in both versions or appropriate descriptions of the searches.
- **IC3**: The study is a fully refereed paper at either a conference or journal, and.
- **IC4**: The study is written in English.

Our exclusion criteria (EC) are listed next, and items that are self-explanatory are not detailed further:

- **EC1**: The study is duplicated (e.g., returned by different databases),
- **EC2**: The original study is not an SLR. Here both the original and the update must be SLRs, so to provide a basis for comparison (e.g. search string used),
- **EC3**: The updated study is a systematic mapping study or a survey. The 3PDF is a decision tool to be applied to SLR updates only, so within the context of this research we maintained the same focus,
- **EC4**: The updated study is a previous version of a more recent update. Here we applied a similar approach as to the one used in SLRs with regard to primary studies: whenever there are different versions of the same primary study, the one to use is the most complete description. Within our context, the latest update would be assumed to be the most complete,

- **EC5**: The study is not a fully refereed paper at either a conference or journal. Here we are neither interested in grey literature nor very brief descriptions such as those in short papers, and.
- **EC6**: The study is written in a language other than English.

## Identifying Search Terms to Use

The first step, prior to carrying out the database searches on several different databases, was to identify the terms to use to build our search string. To do so, the third author carried out a manual search in the proceedings of two conferences – ESEM (International Symposium on Empirical Software Engineering and Measurement) and EASE (International Conference on Evaluation and Assessment in Software Engineering), for the period 2006 to 2018, based upon titles, keywords and abstracts. These conferences were selected for having a clear focus on Evidence-based Software Engineering, and for including secondary studies as some of the topics of interest. We limited the manual search to these two conferences since the objective was to identify the terms for the database search, and not to find studies to include. A total of 794 studies were identified (EASE: 233 and ESEM: 561), and four SE updates were found (three from EASE and one from ESEM proceedings, respectively). The selection was discussed with the first author, and a total agreement was reached. Note that these four studies were also retrieved via our database searches (see Table 3 - S5, S9, S13, and S27), the purpose of our manual search was solely to identify the terms for the database search, and not to find studies to include. All the included studies come from the database searches. During the manual search, terms and synonyms from titles, abstracts and keywords were derived to use later in the database search. The main terms extracted from the manually included studies are shown in Table 1.

The terms in Table 1 led to the following search string:

> *("systematic review" **OR** "systematic literature review" **OR** "systematic mapping")*
> ***AND** (update **OR** extended **OR** extending)*

We used the Boolean **OR** to incorporate synonyms or alternative words or abbreviations and finally, use the Boolean **AND** to link the major terms. The final string was:

> *("systematic literature review" **OR** "systematic review" **OR** "systematic study" **OR** "systematic analysis" **OR** "mapping study" **OR** "systematic mapping" **OR** "mapping review") **AND** (update **OR** updating **OR** updated **OR** extended **OR** extending **OR** extend **OR** extension)*

Note that we used the word mapping to also cater for any possible paper(s) that were genuine SLRs but perhaps labelled themselves as mapping studies.

**Table 1: Terms derived from title, abstract and keywords found in studies from manual search.**

| Study | Conference | Main terms | Keywords |
|-------|-----------|-----------|----------|
| An extended systematic review of software process improvement in small and medium web companies | EASE 2011 | Extended Systematic Review | Software Process Improvement Systematic Review Small Web Companies Medium Web Companies |
| Motivation in software engineering: a systematic review update | EASE 2011 | Update Systematic Review Systematic Literature Review | Motivation Software Engineering Software Engineer Characteristics Personality Systematic Literature Review |
| Cross- vs. Within-Company Cost Estimation Studies Revisited: An Extended Systematic Review | EASE 2014 | Systematic Review Extended Extending | Cost Estimation Models Cross-company Data Within-company Data Estimation accuracy Systematic Review |
| Replication of Empirical Studies in Software Engineering: An Update of a Systematic Mapping Study | ESEM 2015 | Systematic Mapping Update | Replication Empirical Software Engineering Systematic Mapping Study |

**Databases used for searches**

The databases selected for searching were: ACM Digital Library; IEEE Xplore; ScienceDirect; Scopus and Springer Link. These databases were selected for being databases recommended by previous SLRs as the adequate online databases to use (e.g. Maplesden et al. [21], and Zhang et al. [36]). Note that there are other databases (e.g. EI compendex, Wiley Interscience (Wiley Online), Inspec and Kluwer) that were not selected due to the high degree of overlap between their results and those for the selected databases [21].

**Table 2: Details on the search strategy.**

| Source | Studies Retrieved |
|--------|-------------------|
| Science Direct | 722 |
| Springer Link | 241 |
| IEEE Xplore | 223 |
| Scopus | 640 |
| ACM | 199 |
| Sub-Total | 2025 |
| Duplicates | 356 |
| **TOTAL** | **1158** |

All 1158 potentially relevant primary studies were stored using the JabRef software, an open-source bibliography reference manager. We used the export feature available in many electronic databases to automatically export the details of all potential primary studies (i.e., title, author(s), abstract, keywords, year of publication and the name of the data source) to JabRef. Information from databases that did not support exporting to JabRef was manually recorded by the third author. The third author was also the one dealing with duplicates. Finally, the selection of studies was carried out using two phases, as follows.

- **Phase 1:** the first and third authors independently applied the inclusion & exclusion criteria to screen 1065 and 1158 studies, respectively. The strategy used by both was different. The first author clustered, with the JabRef tool support, studies using searches based on terms, e.g., studies that contained the *"systematic review"* term in its title. Subsequently, in this subset of studies, she identified studies that contained, e.g., terms such as *"update"* or *"extended"*. In each subset of studies, the selection criteria were applied by reading the titles and abstracts. The third author sequentially applied the inclusion/exclusion criteria reading the titles/abstracts of all studies organized in the JabRef tool. The individual results were compared and resolved by consensus; a subset of 27 studies was chosen for Phase 2 (detailed next).

- **Phase 2:** the full text of the 27 primary studies included in the preliminary selection (see Table 3) was evaluated independently by the first and third authors to decide whether the study should be included or excluded. The few disagreements regarding the eligibility of studies were also resolved by consensus. The confusion matrix corresponding to the selection is given in Figure 2. It clearly shows that there was a very high overlap between the two authors as to which papers were judged as relevant (23 studies). This meant that Cohen's kappa turned out to be negative (-0.059), suggesting that the two authors' agreement strength is worse than what would be expected by chance alone. A possible explanation for the results could be the detailed level of screening applied in Phase 1, leaving for Phase 2 just a small number of false positives. A total of 20 studies were selected, and seven excluded. The reasons for exclusions are shown in Table 4, and these excluded studies' references are listed in Appendix A.

**Table 3: Studies analysed during Phase 2.**

| ID | Study Title | Study Selection | |
|---|---|---|---|
| | | Initial | Final |
| **S1** | A Survey on Software Release Planning Models | I | I |
| **S2** | A Systematic Literature Review of Techniques and Metrics to Reduce the Cost of Mutation Testing | I | I |
| **S3** | A systematic literature review on cloud computing adoption and migration | I | I |
| **S4** | An extended systematic literature review on provision of evidence for safety certification | I | I |
| **S5** | An extended systematic review of software process improvement in small and medium web companies | I | I |
| **S6** | An update to the SLR of empirical evidence of the impacts and outcomes of computer games and serious games | I | I |
| **S7** | Cloud service evaluation method-based Multi-Criteria Decision-Making: A systematic literature review | I | I |
| **S8** | Critical success factors for ERP system implementation projects: An update of literature reviews | I | I |
| **S9** | Cross vs. within-company cost estimation studies revisited: An extended systematic review | I | I |
| **S10** | Effort Estimation in Agile Software Development: An Updated Review | I | I |
| **S11** | Knowledge management practices in GSD – a systematic literature review update | I | I |
| **S12** | Maintainability prediction of relational database-driven applications: A systematic review | I | I |
| **S13** | Motivation in software engineering: a systematic review update | I | I |
| **S14** | Open-Source Development Tools for Domain-Specific Modeling: Results from a SLR | I | I |
| **S15** | Process simulation for software engineering education | I | I |
| **S16** | Revisiting software ecosystems Research: A longitudinal literature study | I | I |
| **S17** | State-of-the-Art: A Systematic Literature Review on Agile Information Systems Development | I | I |
| **S18** | Systematic literature review on agile practices in global software development | I | I |
| **S19** | Updating a systematic review about selection of software requirements elicitation techniques | I | I |
| **S20** | What is the Further Evidence about UML? - A Systematic Literature Review | I | I |
| **S21** | How personality traits are interrelated with team climate and team performance in SE? A preliminary study | I | E |
| **S22** | Is There a Place for Qualitative Studies when Identifying Effort Predictors? A Case in Web Effort Estimation | I | E |
| **S23** | Obtaining well-founded practices about elicitation techniques by means of an update of a previous SR | I | E |
| **S24** | SOA Reuse: Systematic Literature Review Updating and Research Directions | I | E |
| **S25** | Software process simulation modeling: An extended systematic review | I | E |
| **S26** | Software process simulation modeling: preliminary results from an updated systematic review | I | E |
| **S27** | Replication of Empirical Studies in Software Engineering: An Update of a Systematic Mapping Study | I | E |

| | | Evaluator B | | | |
|---|---|---|---|---|---|
| | | **0** | **1** | | |
| **Evaluator A** | **0** | 0 | 3 | 3 | 12% |
| | **1** | 1 | 23 | 24 | 92% |
| | | 0 | 33 | 26 | |
| | | 4% | 100% | | |

**Figure 2: Confusion matrix for Phase 2.**

**Table 4: Studies excluded during Phase 2 and reasons for the removal.**

| ID/ Ref | Study's Title | Reason for exclusion |
|---|---|---|
| **S21**/[1] | How personality traits are interrelated with team climate and team performance in SE? A preliminary study | **EC3:** It is not an SLR updated. It is a data survey |
| **S22**/[57] | Is There a Place for Qualitative Studies when Identifying Effort Predictors? A Case in Web Effort Estimation | **EC2:** The original study is not an SLR. It is a traditional review |
| **S23**/[56] | Obtaining well-founded practices about elicitation techniques by means of an update of a previous SR | **EC4:** It is the oldest version of study S19 [54] |
| **S24**/[37] | SOA Reuse: Systematic Literature Review Updating and Research Directions | **EC2:** The original study is not an SLR. It is an MS |
| **S25**/[35] | Software process simulation modeling: An extended systematic review | **EC4:** It is one of the original studies of S15 [68] |
| **S26**/[15] | Software process simulation modeling: preliminary results from an updated systematic review | **EC4:** It is one of the original studies of S15 [68] |
| **S27**/[61] | Replication of Empirical Studies in Software Engineering: An Update of a Systematic Mapping Study | **EC2:** The original study is not an SLR. It is an MS |

A summary of the entire process, from selecting search terms to identifying the SE SLR updates, is shown in Figure 3.

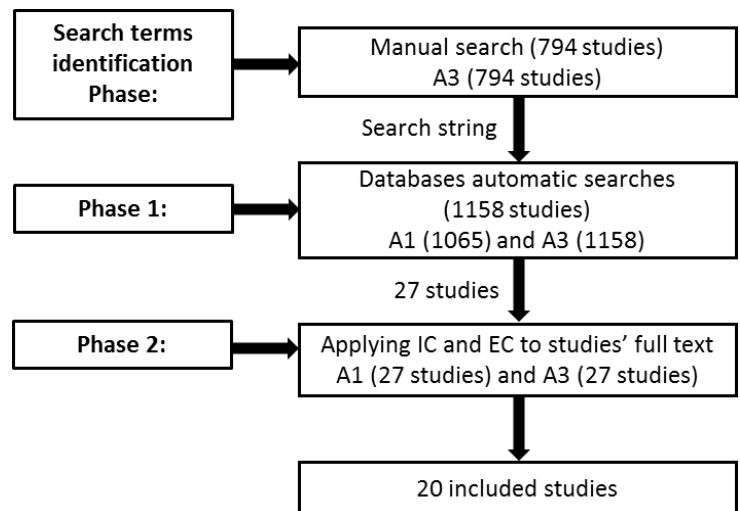

**Figure 3: Summary of the process used to identify search terms and SLR updates.**

Note that in three cases (S9, S15 and S18 in Table 5), the original SLR (OS) had already been revisited at least twice. Under such circumstances, we chose to compare the latest SLR update with its previous most recent version (e.g. S15 is compared to S15_OSc). The only exception is S9, which is compared to S9_OSa (instead of S9_OSb). This was done because, although S9_OSa was not the most recent version before S9, it was the version that presented results for the same subset of research questions discussed in S9. In fact, S9_OSa and S9_OSb complement each other and represent the same SLR.

**Answering the Research Questions**

**RQ1 -** Does the published SLR still address a current question?

Table 5 provides some details about the SE SLR updates that were included herein. The oldest SE SLR update dates back to 2008 (S19), followed by two published in 2011 (S5 and S13). Perhaps one could argue that topics published many years ago would not be meaningful to SE anymore. However, the chosen topics – requirements elicitation techniques, motivation in SE, and software process improvement, are still relevant nowadays and are explicit topics of some SE conferences and journals. Furthermore, all the other SLR updates were published quite recently. Thus, our joint expert judgement is that all the 20 SE SLR updates presented in this section addressed current important SE questions.

**Table 5: Details on types of search and where searches were done.**

| | Updated SLR | | | | Original/Previous SLR | | | |
|---|---|---|---|---|---|---|---|---|
| **ID** | **Research Topic** | **Year** | **Search span** | **Timespan bt. SLRs' publication dates** | **Original SLR (OS)** | **Search span** | **Year** | **Cited by Google Scholar/avg. Citations per year** |
| **S1** [20] | Software Release Planning | 2017 | 2009 – 2015 | 7 years | S1_OS [49] | open – 2008 | 2010 | **158/17.6** |
| **S2** [6] | Mutation Testing | 2019 | 2018 – 2019 | 1 year | S2_OS [28] | 1988 – 2018 | 2018 | 4/4 |
| **S3** [2] | Cloud Computing | 2016 | 2015 – 2016 | 0 year | S3_OS [3] | 2005 – 2015 | 2016 | 1/0.33 |
| **S4** [69] | Safety Certification | 2014 | open – 2012 | 1 year | S4_OS [71] | open – 2012 | 2013 | 25/4.17 |
| **S5** [51] | Software Process Improvement (SPI) | 2011 | 2008 – 2010 | 2 years | S5_OS [34] | open – 2007 | 2009 | 51/5.1 |
| **S6** [24] | Computer Games and Serious Games | 2016 | 2009 – 2014 | 4 years | S6_OS [72] | 2004 – 2009 | 2012 | **878/125.43** |
| **S7** [30] | Cloud Computing | 2018 | 2006 – 2016 | 5 years | S7_OS [31] | 2007 – 2012 | 2013 | 10/1.7 |
| **S8** [18] | Enterprise Resource Planning system | 2015 | 2010 – 2013 | 3 years | S8_OS [17]** | open – 2010 | 2012 | 21/5.25 |
| **S9** [27] | Cost Estimation | 2014 | 2006 – 2013 | 7 years | S9_OSa [7] | 1990 – 2006 | 2006 | ~~111/8.54~~ |
| | | | | | S9_OSb [9] | 1990 – 2006 | 2007 | **338/28.17** |
| **S10** [25] | Effort Estimation | 2018 | 2014 – 2017 | 4 years | S10_OS [53] | 2001 – 2013 | 2014 | **113/22.6** |
| **S11** [60] | Knowledge Management (GSD*) | 2015 | 2012 – 2014 | 3 years | S11_OS [63] | 2008 – 2012 | 2012 | 00/0 |
| **S12** [47] | Maintainability Prediction | 2012 | 2008 – 2010 | 3 years | S12_OS [46] | 1985 – 2008 | 2009 | **214/21.4** |
| **S13** [4] | Motivation in SE | 2011 | 2006 – 2010 | 3 years | S13_OS [64] | 1980 – 2006 | 2008 | **458/41.64** |
| **S14** [14] | Open-Source Development | 2016 | 2005 – 2012 | 0 | S14_OS [70] | 2005 – 2012 | 2016 | 13/4.33 |
| **S15** [68] | Process Simulation | 2015 | open – 2013 | 1 year | S15_OSa [34] | 1998 – 2007 | 2008 | ~~94/8.55~~ |
| | | | | | S15_OSb [36] | 1998 – 2007 | 2010 | ~~53/5.89~~ |
| | | | | | S15_OSc [15] | 2008 – 2012 | 2014 | 04/0.8 |
| **S16** [40] | Software Ecosystem | 2016 | 2007 – 2014 | 3 years | S16_OS [41] | 2007 – 2012 | 2013 | **116/19.33** |
| **S17** [45] | Agile Software Development | 2014 | 2010 – 2013 | 2 years | S17_OS [73] | 2001 – 2010 | 2012 | **797/113.86** |
| **S18** [62] | Agile practices | 2018 | 2010 – 2016 | 8 years | S18_OSa [67] | 1999–2009 | 2011 | ~~133/16.63~~ |
| | | | | | S18_OSb [66] | 1999–2009 | 2010 | **156/17.33** |
| **S19** [51] | Requirements Elicitation Techniques | 2008 | No new search | 3 years | S19_OS [55] | open – 2005 | 2011 | 141/17.63 |
| **S20** [44] | UML | 2017 | 2009 – 2016 | 6 years | S21_OS [19] | open – 2008 | 2011 | **119/14.88** |
| *\* GSD: Global Software Development.      \*\* Both reviews (S8 and S8_OS) were published in a single paper ([18]).* | | | | | | | | |

**RQ2** - Has the SLR had good access or use?

To answer this RQ, we used as cut-off point the same average citation value that has been documented by Garousi and Fernandes [74]. They gathered data from 71,668 SE papers in the pool indexed by Scopus, to identify the top-100 most cited SE papers. They documented that, using the 71,668 papers as basis, the average citation value was 6.82 per paper. Given the large sample size, we believe that it would be reasonable to adopt within the context of our work the same average citation value as a cut-off. Note that all the updates were considered (e.g. S15 had three updates; S9 and S18 had two each). SLR updates with a yearly average equal or above 6.82 were judged as having had good access or use. The average yearly citation is shown in bold in Table 5. Note that some citations in Table 5 are crossed as those SLRs, despite being shown, are not considered within the context of the work presented herein. Only the most recent previous SLRs are considered. Therefore, we have eleven previous SLRs that were judged as presenting good access or use (S1, S6, S9, S10, S12, S13, S16, S17, S18, S19, and S20).

With regard to using the average citation as a surrogate measure for relevance, one of the reasons for doing so was to comply with the same suggested measures proposed in the 3PDF, where citation count and/or website downloads (e.g. as in ResearchGate) were recommended as surrogates for relevance. Another reason was that citation count is a measure that has been used in quite a number of bibliometrics studies in SE, where citations are employed as a measure of research quality and relevance. One example of such use is provided in the same paper abovementioned, by Garousi and Fernandes [74], where they employ citation count as a surrogate for relevance when identifying the top 100 most highly cited SE papers. They also provide a detailed list with numerous bibliometrics studies in SE that have assumed citation count as a surrogate for relevance. Finally, there are also numerous bibliometric studies in other disciplines that also use citation count as surrogate for relevance [42].

**RQ3** - Has the SLR used valid methods and was it well-conducted?
To answer this RQ, the full text of the SLR updates was used to understand how they had been conducted, and to populate Tables 6 to 8.

Table 6 details the SLR method adopted by the 20 SLR updates. The updated versions' protocols were strongly based on the previous version, and we targeted the search strategy, data extraction, and the data synthesis employed to identify whether there were changes between SLR versions. These categories are also the ones suggested in Garner et al. [58].

As can be seen in Table 6, fifteen SLRs (S2, S3, S5, S6, S8, S9, S10, S11, S12, S13, S14, S15, S16, S18 and S19) answered the same RQs of the previous reviews; four (S4, S7, S17 and S20) added a new question or sub-question, and one (S1) discarded some RQs used in the previous version. Furthermore, fifteen SLRs (S2, S3, S5, S8, S9, S10, S11, S12, S13, S14, S16, S17, S18, S19 and S20) reused the same inclusion/exclusion criteria from the original studies; three

(S4, S6 and S15) included an additional exclusion criterion based on publication date to exclude the studies included in the previous version; and two studies (S1 and S7) included additional inclusion/exclusion criteria.

**Table 6: SLR method adopted during the updates.**

| Research Question (RQ) | |
|---|---|
| Answered the same RQ(s) of OS | S2, S3, S5, S6, S8, S9, S10, S11, S12, S13, S14, S15, S16, S18, S19 |
| Added a new RQ or sub-question | S4, S7, S17, S20 |
| Discarded some RQ(s) of OS | S1 |
| **I/E Criteria** | |
| Used the same I/E criteria of OS | S2, S3, S5, S8, S9, S10, S11, S12, S13, S14, S16, S17, S18, S19, S20 |
| Included new I/E criterion based on publication date | S4, S6, S15 |
| Included new I/E criteria | S1, S7 |
| **Search strategy** | |
| Used the same search strategy of OS | S2, S3, S5, S7, S8, S11, S12, S13, S14, S16, S18, S20 |
| Used snowballing only in OS | S19 |
| Used snowballing only in the updated study | S1, S10 |
| Used a subset of database(s) used in OS | S6 |
| Used additional database(s) to those used in OS | S4, S9 |
| Changed the search string used in OS | S15 |
| Changed the search strategy used in OS | S17 |
| **Quality assessment (QA)** | |
| Used the same quality criteria of OS | S3, S4, S5, S6, S9, S10, S11, S12, S13, S14, S15, S16 |
| Defined additional criteria for quality assessment | S18 |
| Did not apply QA to the updated version | S1, S20 |
| Did not apply QA to both versions | S2, S7, S8, S17, S19 |
| **Data extraction** | |
| Used the same data extraction form of OS | S2, S3, S4, S5, S6, S8, S9, S10, S11, S12, S13, S14, S15, S16, S18, S19, S20 |
| Used partially the data extraction form of OS | S1, S7, S17 |
| **Data synthesis** | |
| Integrated data in OS' results | S2, S3, S5, S11 |
| Compare results with those from OS | S1, S4, S6, S7, S8, S9, S10, S12, S13, S15, S16, S17, S18, S19, S20 |
| Included different results from OS | S14 |
| **Team** | |
| The same team of OS | S3, S5, S8 |
| Involved at least one member of OS' team | S2, S4, S6, S7, S9, S12, S14, S15, S16, S19 |
| Created a new team | S1, S10, S11, S13, S17, S18, S20 |

Concerning the search strategy, it can be seen that twelve SLRs (S2, S3, S5, S7, S8, S11, S12, S13, S14, S16, S18 and S20) used the same search strategy as in the previous versions. S9 did not include snowballing despite it being used in the previous version; conversely, S1 and S10 only used snowballing in the updated study. S6 used a subset of the databases used in the previous version, and S4 and S9 used additional databases to the ones used in the previous versions. Finally, S15 changed the search string and S17 changed the search strategy. Further details relating to the search strategy employed by each SLR update are shown in Table 7.

Next, in relation to the quality assessment criteria applied to primary studies, Table 6 shows that twelve SLRs (S3, S4, S5, S6, S9, S10, S11, S12, S13, S14, S15 and S16) used the same criteria as in the original/previous studies. One SLR (S18) adapted its criteria by changing some questions to 'yes/no' questions, and seven SLRs did not apply a quality assessment (S2, S7, S8, S17 and S19 did not apply the quality assessment to both original/previous and current study; S1 and S20 did not apply a quality assessment to the updated SLR). We believe that the decision to not employ any quality assessment to selected studies could have had a detrimental effect upon an SLR update's findings, conclusions and even credibility. Therefore, we believe that the seven SLRs that did not employ any quality assessment have, at least with respect to some aspects, not being well conducted.

**Table 7: Details of the search strategy adopted during the updates.**

| ID | Search strategy | | |
| --- | --- | --- | --- |
| | **Type of Search** | **Search string** | **Databases (dbs)** |
| **S1** | SN[*] Back/Forward | DS was not used | DS was not used |
| **S2** | DS + SN[*] Back | The same search string as in the OS | The same dbs as in the OS |
| **S3** | DS + MS | The same search string as in the OS | The same dbs as in the OS |
| **S4** | DS + MS | The same search string as in the OS | New dbs were used [*] |
| **S5** | DS | Based on keywords of the OS[*] | The same dbs as in the OS |
| **S6** | DS | The same search string as in the OS | A subset of those dbs used in the OS |
| **S7** | DS | Based on keywords of the OS | The same dbs as in the OS |
| **S8** | DS | The same search string as in the OS | The same dbs as in the OS |
| **S9** | DS + SN[*] Back/Forward | The same search string as in the OS | New dbs were used [*] |
| **S10** | SN[*] Forward | DS was not used | DS was not used |
| **S11** | DS | The same search string in the OS | The same dbs as in the OS |
| **S12** | DS | The same search string in the OS | The same dbs as in the OS |
| **S13** | DS + MS | The same search string as in the OS | The same dbs as in the OS |
| **S14** | DS + SN[*] Back | The same search string as in the OS | The same dbs as in the OS |
| **S15** | MS + DS[*] | A new search string was created | New dbs were used [*] |
| **S16** | DS | The same search string as in the OS | The same dbs as in the OS |
| **S17** | MS | DS was not used | DS was not used |
| **S18** | DS | The same search string as in the OS | The same dbs as in the OS |
| **S19** | OS[**]: DS + SN Backward | No new search | No new search |
| **S20** | DS + SN Forward | The same search string as in the OS | New dbs were added |
| OS – Original Study; DS – Database Search; MS – Manual Search; SN – Snowballing; * – Not used in the OS. ** Concerning S20, there was no new search carried out. S20 included 27 additional studies that were not included in its previous review, as they were difficult to locate. | | | |

Except for S1, S7 and S17, all SLRs adopted the same data extraction form used in the original/previous review. In relation to data synthesis, only four studies (S2, S3, S5 and S11) integrated the results obtained from the inclusion of new studies with the previous SLR results. Fifteen SLRs (S1, S4, S6, S7, S8, S9, S10, S12, S13, S15, S16, S17, S18, S19 and S20) presented the results separately from the previous ones, and S14 included the results only.

Finally, ten SLRs (S2, S4, S6, S7, S9, S12, S14, S15, S16 and S19) involved at least one member of the previous review; seven SLRs (S1, S10, S11, S13, S17, S18 and S20) were performed by a different team, and three SLRs (S3, S5 and S8) were updated by the same team.

Overall, except for the seven SLRs that did not employ any quality assessment of evidence (S1, S2, S7, S8, S17, S19, and S20), all the other 13 SLRs were judged as using valid methods and being well conducted (S3, S4, S5, S6, S9, S10, S11, S12, S13, S14, S15, S16, and S18).

**Table 8: Summary of the SLR method adopted during the updates.**

| ID | RQ | I/E Criteria | Search Strategy | Quality assessment | Data extraction | TOTAL |
|-----|-----|-----|-----|-----|-----|-----|
| S1 | ■ | ■ | ■ | - | ■ | 4 |
| S2 | □ | □ | □ | - | □ | 0 |
| S3 | □ | □ | □ | □ | □ | 0 |
| S4 | ■ | ■ | ■ | □ | □ | 3 |
| S5 | □ | □ | □ | □ | □ | 0 |
| S6 | □ | ■ | ■ | □ | □ | 2 |
| S7 | ■ | ■ | □ | - | ■ | 3 |
| S8 | □ | □ | □ | - | □ | 0 |
| S9 | □ | □ | ■ | □ | □ | 1 |
| S10 | □ | □ | □ | □ | □ | 0 |
| S11 | □ | □ | □ | □ | □ | 0 |
| S12 | □ | □ | □ | □ | □ | 0 |
| S13 | □ | □ | □ | □ | □ | 0 |
| S14 | □ | □ | □ | □ | □ | 0 |
| S15 | □ | ■ | ■ | □ | □ | 2 |
| S16 | □ | □ | □ | □ | □ | 0 |
| S17 | ■ | □ | ■ | - | ■ | 3 |
| S18 | □ | □ | □ | ■ | □ | 1 |
| S19 | □ | □ | - | - | □ | 0 |
| S20 | ■ | □ | □ | - | □ | 1 |

*Legend: □ Same as in the previous study. ■ Changed in relation to the previous study. - no search/quality assessment*

**RQ4** - Are there any new relevant methods?

To answer this RQ, the full text of the SLR updates was used to check whether they presented differences from the original versions in the way they searched for evidence, did the synthesis of evidence, whether they included other types of study designs (e.g. qualitative research), and new statistical techniques or improvements to existing techniques used for meta-analysis. This information is available in Tables 6 to 8. A total of nine SLR updates (S1, S4, S6, S7, S9, S15, S17, S18, and S20) have changed some aspect of the method when compared to their previous SLRs' methods. The largest changes were with regard to the search strategies used (six out of nine studies), and research questions employed (five out of nine studies). Conversely, 11 SLR updates did not change their methods, when compared to their previous SLRs (S2, S3, S5, S8, S10, S11, S12, S13, S14, S16, and S19).

**RQ5** - Are there any new studies, or new information?

Table 5 shows that the only SLR update that did not carry out new searches on some additional time span was S19. Therefore, all the SLR updates, except for S19, had new studies. We used S19's full-text to check whether there were some new information/conclusions provided (Evidence is provided in Table C.1 in Appendix C); however, none was found.

**RQ6** - Will the adoption of new methods change the findings, conclusions or credibility?

This RQ only relates to the nine SLR updates (S1, S4, S6, S7, S9, S15, S17, S18, and S20) that have changed some aspect of the method, when compared to their previous SLRs' methods. We focused upon the evidence provided in Tables 6 to 8 as the basis to answer this RQ. These tables show that S1 and S20 did not apply any quality assessment to selecting new studies. Thus, the credibility of their findings could have been compromised. Furthermore, SLR updates S4, S7, S17, and S20 added a new RQ or sub-question, which can be seen as contributing to increasing the range of findings presented. Therefore, we believe that five SLR updates (S1, S4, S7, S17, and S20) had changes in the findings, conclusions or credibility as the result of changes in the employed methods.

**RQ7** - Will the inclusion of new studies/information/data change findings, conclusions or credibility?

To answer this RQ, the full text of both the original SLRs and their updates were used. We provide a detailed table (Table C.1 in Appendix C) reporting the evidence we used from the SLR updates to decide whether findings/conclusions/credibility changed when compared to previous versions. A total of 15 SLR updates (S1, S2, S4, S6, S7, S8, S9, S10, S11, S12, S14, S16, S17, S18, and S20) have changed their findings, conclusions or credibility as the result of the inclusion of new studies/information/data. Only five SLR updates (S3, S5, S13, S15, and S19) did not change their findings, conclusions or credibility as the result of including new studies/information/data.

All the research questions can also be answered using a 'YES'/'NO' associated with each SLR update. Therefore, a summary for the 20 SLR updates, arranged by RQs, is shown in Table 9, below. The table is used later in Section 5 when applying the 3PDF Framework to each of the 20 SLR updates described in this section. An observation is that no single study provided a 'YES' answers to all RQs.

**Table 9: Summary of studies according to RQs.**

| ID | RQ1 | RQ2 | RQ3 | RQ4 | RQ5 | RQ6 | RQ7 |
|-----|-----|-----|-----|-----|-----|-----|-----|
| S1 | YES | YES | NO | YES | YES | YES | YES |
| S2 | YES | NO | NO | NO | YES | NO | YES |
| S3 | YES | NO | YES | NO | YES | NO | NO |
| S4 | YES | NO | YES | YES | YES | YES | YES |
| S5 | YES | NO | YES | NO | YES | NO | NO |
| S6 | YES | YES | YES | YES | YES | NO | YES |
| S7 | YES | NO | NO | YES | YES | YES | YES |
| S8 | YES | NO | NO | NO | YES | NO | YES |
| S9 | YES | YES | YES | YES | YES | NO | YES |
| S10 | YES | YES | YES | NO | YES | NO | YES |
| S11 | YES | NO | YES | NO | YES | NO | YES |
| S12 | YES | YES | YES | NO | YES | NO | YES |
| S13 | YES | YES | YES | NO | YES | NO | NO |
| S14 | YES | NO | YES | NO | YES | NO | YES |
| S15 | YES | NO | YES | YES | YES | NO | NO |
| S16 | YES | YES | YES | NO | YES | NO | YES |
| S17 | YES | YES | NO | YES | YES | YES | YES |
| S18 | YES | YES | YES | YES | YES | NO | YES |
| S19 | YES | YES | NO | NO | NO | NO | NO |
| S20 | YES | YES | NO | YES | YES | YES | YES |

## 4 A survey involving authors of SLR updates in SE

This section describes a survey aimed to obtain feedback from the authors of the SLR updates presented in Section 3 concerning the suitability of using the 3PDF in an SE context.

The objective was to contact the 74 authors (excluding authors of this paper) of the 20 identified SLR updates by email. Unfortunately, four of them were not reachable under the publication e-mail address anymore. Hence, we invited 70 authors to answer the survey, which was instrumented as an online questionnaire implemented in Google forms (see Appendix B). The authors had 15 days to answer the survey, and some had an additional 15-day period. A total of nine responses were received, resulting in an overall response rate of 12.86% (9/70).

Nevertheless, as suggested for survey research [52], we focused on the target population that would be the most appropriate to provide accurate answers, instead of primarily searching for high response rates. Moreover, despite the low number of respondents, the response rate is similar to that documented in other surveys in SE [32], and common within the SE field [5]. In addition, the nine respondents were co-authors of eight of the 20 SLR updates, and some provided additional expert feedback on the 3PDF to support our analyses and the discussion on the suitability of using the framework within an SE context.

The questionnaire objectively focused on obtaining feedback on the relevance of each of the 3PDF questions (Steps 1.a-3.c in Figure 1) and respondents' agreement with the decision drivers (the arrows in Figure 1). The feedback on the questions and decision drivers is shown in Table 10 and in Table 11, respectively.

**Table 10: Relevance assessment (# respondents) for each of the 3PDF questions.**

| Step Question | 1.a | 1.b | 1.c | 2.a | 2.b | 3.a | 3.b |
|---|---|---|---|---|---|---|---|
| *RELEVANT* | 8 | 7 | 9 | 9 | 8 | 7 | 6 |
| *NOT relevant* | 1 | 2 | 0 | 0 | 1 | 2 | 3 |

It is possible to observe in Table 10 that there was an overall strong agreement on the relevance of the 3PDF questions. Indeed, only three participants (hereafter referred to as A, B, and C) disagreed on the relevance of at least one of the questions. Moreover, only three questions were considered as *NOT relevant* by more than one participant: 1.b, 3.a, and 3.b. We focus the discussion on these.

Regarding question 1.b, which relates to assessing whether an SLR to be updated had good access or use (i.e., if it had an impact on research and/or practice), respondents A and B considered this question as *NOT relevant*. Respondent A mentioned that *"I think it is possible that an SLR that has received a small number of citations and/or has not had a significant impact on practice to become relevant to update"*, while respondent B justified that *"the question might be relevant if considered in an appropriate time frame"*. Nevertheless, we agree with the remaining experts on the relevance of this question as it enables mitigating the risk of investing effort in an SLR update that might not have any impact on research and/or practice. However, we are aware that, as stated by respondent D, *"Relevance to practice might be hard to deduct"*. Also sceptical with regard to question 1.b, Respondent E mentioned that *"While the question is relevant, an SLR might need to be updated even when the answer to 1.b is No, depending on the reasons for the low access or use. E.g., if it is low because of limitations in the original SLR, an update would make sense"*. The 3PDF Framework includes a question that deals specifically with what Respondent E is highlighting – question 2a (Are there any new relevant methods?). A flawed SLR should not be updated (*e.g.*, an SLR that did not use valid methods or was not well conducted), and the 3PDF Framework's question 1.c caters for this. The majority of respondents agree that the 3PDF's question 1.b properly emphasizes that research efforts should be directed towards generating impactful results.

Questions 3.a and 3.b respectively concern the assessment of whether the adoption of new methods or the inclusion of new studies will change the SLR's findings or credibility. Question 3.a was considered as *NOT relevant* by participants A and B, while question 3.b was considered as *NOT relevant* by participants A, B, and C. Respondent A stated that *"The answer to this question may be the outcome of the updated study!"*. Respondent B believes that the findings and credibility would always be affected "I would always answer these questions YES, no matter what." Respondent C did not justify her/his answer. In line with respondent A's statement, respondent E, while still considering the questions relevant, mentioned that *"In our case, the update strengthened our results and conclusions. They remain practically the same, so the original ones were valid. This also raised the question of the extent to which the update (and the required effort) was really needed, as the overall picture of the area studied didn't change. Anyway, this can't be ascertained without an update"*. However, as can be seen in Figure 1, the 3PDF relaxes this condition and requires only one of those questions to be answered with a *YES* or *MAYBE*. Indeed, answering question 3.a requires evaluating the methods employed in the original SLR and comparing them with the intended new methods to use in the SLR update to assess potential effects; while answering question 3.b requires searching for new studies and conducting an informal evaluation of the effects based on reading titles and abstracts. Nevertheless, we agree with the majority of respondents and believe that the upfront effect assessment that should be conducted in steps 3.a and 3.b is beneficial as it avoids investing effort that will not reflect positively on the evolution of the SE evidence corpus.

Table 11 shows the results on the decision drivers. Again, a strong agreement can be observed. Respondents A and B, who were sceptical on some of the 3PDF questions, did also not agree on the decision drivers. Respondent A did not agree with Steps 1 and 3, while respondent B did not agree with Steps 2 and 3. Additionally, respondent E did not agree with the decision driver for Step 1. The reasons are directly related to the previously mentioned statements regarding the questions. Respondent A believes that low access SLRs might become relevant (Step 1) and that instead of an effect assessment, the SLR should be updated (Step 3). Respondent B believes that most SLRs are restricted to a certain time frame (Step 2) and that the findings and credibility would always be affected (Step 3). Respondent E considers that SLRs that have low access due to limitations should still be updated (Step 1). Note that methodological limitations (not flawed methodologies) are catered for by the 3PDF's question 2.a. Furthermore, if the use of a flawed methodology has had an effect upon an SLR's access, a totally new SLR should be the solution – one that employs proper methods and research questions. Finally, as previously stated, most respondents agreed that the 3PDF decision drivers are relevant to be employed to decide whether an SLR is suitable for an update.

It is important to note that two of the three respondents who criticized some of the questions in the 3PDF are part of the subset of authors whose SLR updates would not have been assessed as suitable for an update. This issue is also discussed in Section 6 – Threats to validity.

**Table 11: Feedback (# respondents) on the agreement with the 3PDF decision drivers.**

| Decision Driver | ALL questions in Step 1 MUST have a "YES" answer, before moving on to Step 2, when considering ONLY the questions in Step 1 that you judged as RELEVANT? | AT LEAST ONE of the questions in Step 2 MUST have a "YES" answer before moving on to Step 3, when considering ONLY the questions in Step 2 that you judged as RELEVANT? | AT LEAST ONE of the questions in Step 3 MUST have a "YES" answer, when considering ONLY the questions in Step 3 that you judged as RELEVANT? |
|---|---|---|---|
| *Agree* | 7 | 8 | 7 |
| *Disagree* | 2 | 1 | 2 |

As a result of the survey response analyses, we put forward that the 3PDF was deemed suitable for use within the SE context. Respondent E, for example, argues that *"The framework is relevant and was somehow implicitly used in our decision to update the SLR"*. Respondent F emphasizes that *"Although some aspects of the framework might seem obvious to some people, I think that it can help researchers to decide upon a possible update"*. Indeed, making the 3PDF available to the SE community may help researchers decide on whether to conduct an SLR update and, in case of deciding for an update, to explicitly provide arguments on the rationale employed to make such decision. Moreover, it may serve as guidance for reviewing the relevance of SLR updates. The next section presents an assessment of the suitability for an update of each of the 20 SE SLR updates described in Section 3.

## 5 Applying the 3PDF to assess the need for updating SLRs

Table 12 shows the results from applying the 3PDF (Figure 1) to each of the 20 SE SLR updates. This assessment was based upon the summary table already provided in Section 3 – Table 9.

Also note that the research questions are directly related to the seven steps part of the 3PDF, as outlined in Section 3:

RQ1 – Step 1.a: Does the published SLR still address a current question?

RQ2 - Step 1.b: Has the SLR had good access or use?

RQ3 - Step 1.c: Has the SLR used valid methods and was well-conducted?

RQ4 - Step 2.a: Are there any new relevant methods?

RQ5 - Step 2.b: Are there any new studies, or new information?

RQ6 - Step 3.a: Will the adoption of new methods change the findings, conclusions or credibility?

RQ7 - Step 3.b: Will the inclusion of new studies/information/data change findings, conclusions or credibility?

**Table 12: Application of the updated 3PDF to previous SLR updates in SE.**

| SID | 1.a | 1.b | 1.c | 2.a | 2.b | 3.a | 3.b | Recommendation |
|-----|-----|-----|-----|-----|-----|-----|-----|----------------|
| S1 | YES | YES | NO | - | - | - | - | No: Update not needed |
| S2 | YES | NO | NO | - | - | - | - | No: Update not needed |
| S3 | YES | NO | YES | - | - | - | - | No: Update not needed |
| S4 | YES | NO | YES | - | - | - | - | No: Update not needed |
| S5 | YES | NO | YES | - | - | - | - | No: Update not needed |
| S6 | YES | YES | YES | YES | YES | NO | YES | **Yes: Prepare update** |
| S7 | YES | NO | NO | - | - | - | - | No: Update not needed |
| S8 | YES | NO | NO | - | - | - | - | No: Update not needed |
| S9 | YES | YES | YES | YES | YES | NO | YES | **Yes: Prepare update** |
| S10 | YES | YES | YES | NO | YES | - | YES | **Yes: Prepare update** |
| S11 | YES | NO | YES | - | - | - | - | No: Update not needed |
| S12 | YES | YES | YES | NO | YES | - | YES | **Yes: Prepare update** |
| S13 | YES | YES | YES | NO | YES | - | NO | No: SLR up to date |
| S14 | YES | NO | YES | - | - | - | - | No: Update not needed |
| S15 | YES | NO | YES | - | - | - | - | No: Update not needed |
| S16 | YES | YES | YES | NO | YES | - | YES | **Yes: Prepare update** |
| S17 | YES | YES | NO | - | - | - | - | No: Update not needed |
| S18 | YES | YES | YES | YES | YES | NO | YES | **Yes: Prepare update** |
| S19 | YES | YES | NO | - | - | - | - | No: Update not needed |
| S20 | YES | YES | NO | - | - | - | - | No: Update not needed |

Step 1.a Does the published SLR still address a current question?
Step 1.b Has the SLR had good access or use?
Step 1.c Has the SLR used valid methods and was well-conducted?
Step 2.a Are there any new relevant methods?
Step 2.b Are there any new studies, or new information?
Step 3.a Will the adoption of new methods change the findings, conclusions or credibility?
Step 3.b Will the inclusion of new studies/information/data change findings, conclusions or credibility?

The results from applying the 3PDF framework to assess the suitability of an SLR for update showed that only six of the 20 studies would have legitimate grounds for updating a previously conducted SLR (S6, S9, S10, S12, S16, and S18).

Nine original/previous SLRs (See Table 5) had average yearly citations below our cut-off point (S2, S3, S4, S5, S7, S8, S11, S14, and S15), which led to a 'NO' answer to Step 1b. Seven SLR updates did not use any quality criteria to assess studies (S1, S2, S7, S8, S17, S19, and S20), which led to a 'NO' answer for Step 1c. Four SLR updates did not use new methods, which led to a 'NO' answer for Step 2a. Three SLR updates included new methods; however, the inclusion did not change the studies' findings, conclusions or credibility (S6, S9, and S18). This led to a 'NO' answer for Step 3a. Finally, the inclusion of new studies in S13 did not change the findings, conclusions or credibility; this led to a 'NO' answer for Step 3b. Note that the outcome in Table 12 is based on having updates for comparison. It may be more challenging to answer, for example, Step 3b without having done or at least attempted to do an update. Furthermore, the 3PDF framework may inspire researchers to think carefully about ways to add new value with an update, for example, decide to use a new method to motivate an update.

In essence, based on the 20 studies discussed herein, Step 1b ('NO' answer to 9 original/previous SLRs), followed by step 1c ('NO' answer to seven SLR updates), and 3b ('NO' answer to one SLR update) were the steps determining whether an SLR was suitable for an update or not.

Steps 1b and 1c are regarded in the 3PDF as crucial to deciding whether an SLR is suitable for an update, and hence the need for all answers to be 'YES' to proceed to Step 2. These steps can be summarised as: Are the SLR findings contributing to research and practice? Can we trust the results that are to be provided in the SLR update?

Note that Step 1b is driven by how relevant the original/previous SLR has been to the research community, or to practice, or to both, based on the average yearly citation and/or downloads. The decision as to which topic to select as an SLR can be informed, for example, by a preliminary mapping study on the topic, so to find out existing gaps and also whether such mapping study has had good access or use.

Note that, as shown in Table 5, six of the nine SLR updates associated with the nine original/previous SLRs that presented an average yearly citation below our cut-off point, were published up to only one year after the publication of the original/previous SLRs. Such a short time span may explain the low average yearly citations shown for the original/previous SLRs.

With regard to Step 1c, methodological rigour is expected and should be employed in every SLR, regardless of it is an update of a previously published SLR.

With regard to Step 3b, it addresses the question as to whether the new research findings in the SLR update would make any significant contributions to the state-of-the-art when compared to what was already known via the original/previous SLR. The answer to this step requires some effort as one needs to carry out a few preliminary searches to perform a risk analysis on whether the SLR update's research contribution motivate an update. However, this is still worth doing in order to at least reduce the number of SLR updates that are being published that are not needed, according to the 3PDF framework.

# 6 Threats to validity

## 6.1 Threats related to identifying the SE SLR updates and assessment

The four types of validity threats and corresponding aspects of threats that are discussed in this sub-section (see Table 13) have been identified by Zhou et al. [76] as the most discussed threats and aspects in SE SLRs, based upon data extracted from 316 SE SLRs published within the period 2004 to 2015.

Four types of validity threats have been identified [76]:

**Internal (In):** Unknown factors that may affect the dependent variable. E.g., confounding factors we are unaware of.

**External (Et):** To what extent we can generalise the findings.

**Conclusion (Cn):** To be able to draw correct conclusions regarding the relationships being investigated, e.g. use of an adequate statistical test, use of proper measurement

**Construct (Ct):** Represents to what extent operational measures precisely measure the concepts they claim to measure.

The aspects of validity threats that are discussed herein are the following [76]:

### Table 13: Validity & Aspects of Threats and Treatments.

| Aspects | Treatment |
|---|---|
| Incorrect or incomplete search terms employed in database searches | We carried out a manual search in the proceedings of two conferences – ESEM (International Symposium on Empirical Software Engineering and Measurement) and EASE (International Conference on Evaluation and Assessment in Software Engineering), for the period 2006 to 2018, based upon titles, keywords and abstracts. These conferences were selected for having a clear focus |

| | on Evidence-based Software Engineering, and for including secondary studies as some of the topics of interest. A total of 909 papers were screened. Later, we also added synonyms to search terms, prior to carrying out the database searches. |
|---|---|
| Incomprehensive venues or databases | First, the databases selected for use with the database searches were: ACM Digital Library; IEEE Xplore; ScienceDirect; Scopus and Springer Link. These databases were selected for being databases recommended by previous SLRs as the most adequate online databases to use (e.g. Maplesden et al. [21], and Zhang et al. [36]). Note that there are other databases (e.g. EI compendex, Wiley Interscience (Wiley Online), Inspec and Kluwer) that were not selected due to the high degree of overlap between the results and those for the selected databases [21]. Second, we had access to all the above mentioned databases through our Institutions' subscriptions. |
| Bias in study selection | To identify terms for the search string, the third author carried out a manual search on proceedings and later discussed the selection with the first author, and a full agreement was reached. Both the first and third authors searched independently through reading titles and abstracts for the candidate studies retrieved via database searches and later compared their respective selection until consensus was reached. Later, both the first and third authors looked at the full text of the 27 studies that were selected from the first phase and selected 20 with a high level of agreement. Finally, both authors have experience carrying out a number of SLRs. |
| Misclassification of primary study | With regard to this work, the classifications corresponded to the different steps part of the 3PDF framework, and also represented by this SLR's research questions (RQs). The data needed to answer the RQs was extracted by the third author and always discussed with the first author until consensus was reached. Thus, we had two researchers working through each of the SLR updates, and we do believe that such process mitigated classification biases. |
| Publication bias | Within the context of this SLR, any SE SLR updates identified via our searches were included, regardless of whether they included additional evidence from more recently published primary studies and/or new conclusions. |

| Bias in data extraction | Bias could have been introduced given two co-authors in this paper (first and fourth) also co-authored one of the SLR updates. |
| | The second issue was addressed by forming a studies selection team that included one author (third author) who did not co-author any SLR update, and who was also responsible for extracting the data from the 13 SLR updates that were selected. |

**Incorrect or incomplete search terms employed in database searches**: The search string may include inadequate search terms related to the SLR topic, and/or not include terms that should have been included. Associated with Construct and Internal validity.

**Incomprehensive venues or databases**: Some important databases are not available for not being subscribed by the Institutions to which the authors are affiliated. Associated with Construct validity.

**Bias in study selection**: While searching for candidate studies, researchers select studies subjectively, without completely complying with the inclusion and exclusion criteria to support their decision. Associated with Internal and Conclusion validity.

**Misclassification of primary study**: "Researchers may not consider all the classifications at the beginning of the research process or the primary studies selected by the reviewers are mistakenly classified" [76]. Associated with Internal and Conclusion validity.

**Publication bias**: Bias associated with the problem that positive results are more likely to be published than negative results [7]. Associated with Internal validity.

**Bias in data extraction**: While extracting data, researchers may not totally understand what needs to be extracted, and how the data relates to the research questions. Associated with Internal and Conclusion validity.

## 6.2   Threats related to the survey with SE SLR update authors

We identified some threats throughout the process of planning and conducting the survey. Table 14 lists these potential threats, classified by the survey validity types presented in [38], and explains how they were addressed in our survey. One of the main mitigation strategies was validating the instrument amongst the authors of this paper, before handing it over to the SE SLR update authors.  All four authors of this paper have carried our numerous SLRs and have also participated in SLR updates.

There was an additional threat that relates to a possible bias in the respondents' answers to motivate the SLR update(s) they co-authored. Out of the nine respondents, four co-authored SLR updates that would have been updated if the 3PDF framework had been used; all of them were very supportive of the Framework's questions. Conversely, the only three respondents who were critical of the Framework would not have had their SLR updates selected if applying the 3PDF framework. However, another two respondents, who would also not have had their SLR updates selected if applying the 3PDF framework, were very supportive of the Framework's questions. Therefore, we argue that the evidence presented here cannot be argued to suggest a bias for or against the use of the 3PDF framework.

**Table 14: Threats to Validity and Treatments.**

| Threats | Treatment |
|---|---|
| Face Validity – Bad instrumentation | Revision and evaluation of the questionnaire about the format and formulation of the questions. Questions objectively focused on the 3PDF. Running a pilot study. |
| Content Validity – Inadequate explanation of the constructs | Revision and evaluation of the questionnaire about the format and formulation of the questions. Running a pilot study. Providing a brief explanation on the 3PDF and a link with further details. |
| Criterion Validity – Not surveying the target population. | We identified SE SLR update authors following an explicitly documented and carefully conducted procedure (cf. Section 3). |
| Construct Validity – Inadequate measurement procedures and unreliable results. | We only used frequency counting, which can be safely applied to discrete survey questions concerning the relevance of the 3PDF questions and the agreement with the 3PDF decision drivers. Also, we triangulated the answers with the provided explanations. |
| Reliability – Lack of statistical conclusion validity | This threat strongly depends on the sample size. Unfortunately, while contacting twice the SE SLR update authors we were aware of, our final sample size was still limited. Hence, we focused our results on qualitative analyses and did not make any further claims on conclusion validity. |

## 6.3   Threats related to applying the 3PDF to assessing the SE SLR updates

In relation to validity threats with regard to applying the 3PDF to assessing the SE SLR updates, we believe that the three main validity issues herein relate to whether: i) authors misunderstood the 3PDF framework; ii) bias being introduced given two co-authors of this paper (first and fourth) also co-authored one of the SLR updates; iii) bias being introduced when applying the 3PDF framework to the SLR updates.

With regard to the first issue, the original paper detailing the 3PDF Framework was shared and discussed between all the authors, to solve any doubts/misunderstandings/etc. The 3PDF Framework was also used as a basis for the instrument and corresponding survey, which are detailed in Section 4. None of the survey's respondents mentioned anything related to ambiguity in regard to the survey's instrument, which we believe can also be used as additional evidence showing that the 3PDF Framework was understood by the authors.

In relation to the second issue, there was only one SLR update that has been co-authored by the first and fourth co-authors in this paper (S9). All the summary Tables 5 to 8, and also C.1 in Appendix C were prepared by the third author, and later partially re-checked by the first author (e.g. text in Table C.1, all of Table 8, naming consistency in Table 5, and adding average yearly citation). The content was used to answer the research questions outlined in Section 3, where we also provide details on how each RQ has been addressed. Further, Table 9 provides a summary for all the 20 studies, and the corresponding answers 'YES' or 'NO' can be checked against the summary data shown in Tables 5 to 8, Table C.1, and the explanations provided about how each RQ was answered. Section 3 provides the data needed to assess each of the 20 SLR updates. In fact, Table 9 was directly used in Section 5. Finally, study S9 was assessed by the third author.

Finally, as for the third issue, Section 3 provides research questions that match all the steps part of the 3PDF Framework directly, and Table 9 can be used as input to the assessment shown in Table 12. Thus, we believe that a good quality Section 3 is directly related to how well the assessment was done in Section 5.

Section 3 provides summary tables (Tables 5 to 8, and Table C.1 in Appendix C), and details how the tables were used to answer each RQ. Finally, an overall summary table (Table 9) is also provided, so to enable anyone to check it against the summary data shown in Tables 5 to 8, Table C.1, and the explanations provided under each RQ on how the RQs were answered.

These tables were prepared by the third author based on the full-text for each of the 20 studies and were later on partially re-checked by the first author (who also used the full-text for the included studies). Once the assessment was done, basically via using Table 9 to prepare Table 12, the second and fourth authors evaluated the results to see if anything unusual/mistake could be identified. None were found.

## 7  Implications

The results from applying the 3PDF Framework to assess the suitability for updating the 20 SE SLR updates showed that half of the 14 SLR updates assessed as unnecessary to update were published within a time span of no more than two years from the publication of the original/previous SLR. These are: S2 – mutation testing, S3 – cloud computing, S4 – safety certification, S5 – software process improvement, S14 – open-source development, S15 – process simulation, and S19 - requirements elicitation techniques. None of the SLRs assessed as suitable for updating were published within a time span of no more than two years from the publication of the original/previous SLR. The shortest time span was three years, for SLR updates: S12 (maintainability prediction), S13 (motivation in SE), and S16 (software ecosystem). We believe that one of the several take away messages for the SE research community is that the 'maturation' time between the publication of the original/previous SLR and its update should be longer than two years, although exceptions exist. Such suggestion does not however bypass the need to apply the 3PDF to assess whether an SLR update is genuinely needed. Another take away message is that our results show that many resources are wasted in carrying out updates that did not need to be conducted. This applies to all the 14 SLR updates that were assessed as unsuitable for an update, not only the eight abovementioned.

Furthermore, the results detailed in Section 5 also illustrate that several SLR updates did not employ any quality control when selecting primary studies, which has clear detrimental implications to the quality of the evidence being provided by these SLR updates. This applied to seven SLR updates (S1 - 2017, S2 - 2019, S7 - 2018, S8 - 2015, S17 - 2014, S19- 2008, and S20 - 2017). It is a surprising result given that there are numerous guidelines and literature on how to carry out SLRs in SE, and they all stress the use of quality checklists (e.g. [7][12]). Note that, except for S19, all the other six SLR updates were published in the past five years, thus quite recently. As pointed out in Section 1, there are now hundreds of SE SLRs published. Therefore, another take away message is that we believe that it is important that not only SLR authors, but also reviewers of SLRs submitted to conferences and journals should be familiar with SLR guidelines to avoid the publication of SLRs that are methodologically flawed.

Another important aspect that has been brought to light is that, whenever SLR updates do not include new relevant methods, they should at least provide results that reflect a change in findings, conclusions, or credibility, when compared to the original/previous SLR. The point here is that the effort to carry out an SLR update should be justified by some novelty, in either the methods used or new findings/conclusions/credibility. If nothing has changed, one may as well rely on the findings already presented in the original/previous SLR. This would also apply to newly starting PhD students, i.e., they should not waste any effort to carry out an SLR update if such update is not justified by a novelty in either the methods used or new findings/conclusions/credibility. And the justification for the decision to not update (or to update) an existing SLR can be based upon applying the 3PDF to the original SLR, and additionally supported by the evidence and recommendations provided herein.

This paper has focused on the "When" to update SLRs. However, we also intend to tentatively address the "How" to update SLRs. We have reproduced here Table 8 (shown as Table 15 below). It highlights the six SLR updates that we assessed as suitable for an update, as per the 3PDF framework. Four of these used exactly the same RQs, I/E Criteria, search strategy, quality assessment and data extraction as used in the original/previous SLRs (S10, S12, S16, and S18). S9 used additional databases for the database searches, and S6 used a subset of the databases. Finally, S6 added a new I/E criterion based on publication date. Based on the evidence, we believe that an additional take away message would be to employ the same mechanisms used in the original/previous SLR being updated. This was the case in four of the six SLR updates discussed herein. In addition, we have made recommendations elsewhere as to what search strategies to employ when updating SE SLRs [26], using one of the SLR updates presented herein (S9), and two of its replications. These can complement the suggestions abovementioned regarding "How" to update SLRs in SE.

There is one last comment we would like to make. The research process that we have employed herein could also be seen as a set of Design Science (DS) activities: i) problem analysis, which is characterised by our assessment relating to the suitability for update of SE SLR updates; ii) solution design, which is characterised by using the 3PDF framework; and iii) solution evaluation, which is characterised by our survey. We did not carry out the research detailed herein with DS in mind. However, we believe that it is important to stress that DS would also have been a suitable match for our research process.

**Table 15: Summary of the SLR methods adopted during the updates.**

| ID | RQ | I/E Criteria | Search Strategy | Quality assessment | Data extraction | TOTAL |
|----|----|----|----|----|----|----|
| S1 | ■ | ■ | ■ | - | ■ | 4 |
| S2 | □ | □ | □ | - | □ | 0 |
| S3 | □ | □ | □ | □ | □ | 0 |
| S4 | ■ | ■ | ■ | □ | □ | 3 |
| S5 | □ | □ | □ | □ | □ | 0 |
| S6 | □ | ■ | ■ | □ | □ | 2 |
| S7 | ■ | ■ | □ | - | ■ | 3 |
| S8 | □ | □ | □ | - | □ | 0 |
| S9 | □ | □ | ■ | □ | □ | 1 |
| S10 | □ | □ | □ | □ | □ | 0 |
| S11 | □ | □ | □ | □ | □ | 0 |
| S12 | □ | □ | □ | □ | □ | 0 |
| S13 | □ | □ | □ | □ | □ | 0 |
| S14 | □ | □ | □ | □ | □ | 0 |
| S15 | □ | ■ | ■ | □ | □ | 2 |
| S16 | □ | □ | □ | □ | □ | 0 |
| S17 | ■ | □ | ■ | - | ■ | 3 |
| S18 | □ | □ | □ | ■ | □ | 1 |
| S19 | □ | □ | - | - | □ | 0 |
| S20 | ■ | □ | □ | - | □ | 1 |

*Legend: □ Same as in the previous study. ■ Changed in relation to the previous study. - no search/quality assessment*

# 8 Conclusions

This paper investigates and makes recommendations towards an important aspect relating to the update of SLRs in SE – when to update an SLR. We initially formulated four overall research questions in Section 1. To summarise, ORQ1 is addressed by presenting some proposals for how to decide when to update an SLR in Section 2. Next, ORQ2 is addressed by conducting a systematic search for updates in SE, which is presented in Section 3. ORQ3 is addressed through a survey of authors having published updated SLRs in SE. The survey is presented in Section 4. Finally, ORQ4 is addressed by assessing the retrieved updates of SLRs using the 3PDF framework. The outcome is presented in Section 5.

In more detail we address the challenge of deciding when to update an SLR using three consecutive stages: Stage 1, explained in Section 3, details a two-step process (see Figure 3) used to identify relevant SLR updates in SE, which provide not only the set of 20 SE SLR updates to be employed in Section 5, but also the authors, who were contacted via a survey and invited to provide their feedback about the suitability of a third-party decision framework (3PDF), proposed in [58], to be used by the SE community to assess the suitability of SE SLRs for updating. The survey – our Stage 2, is detailed in Section 4. Nine authors answered the survey, out of the 74 who were contacted twice by email. As Stage 3, we have applied the 3PDF framework to assess the suitability of each of the 20 SE SLR updates for updating (detailed in Section 5).

After applying the 3PDF, results showed that 14 of the 20 SE SLR updates did not need updating. Based on the assessment of the 20 SLR updates, the main decision driver to whether an SLR should be updated or not was Step 1b - the SLR's contribution to research and practice, which was given a 'NO' answer to nine of SLR updates. Another two steps – Steps 1c and 3b, received 'NO' answers to another seven and one SLR updates, respectively.

The results, in our view, suggest that the use of a decision support mechanism (as the 3PDF) to help the SE community decide upon the need for SLR updates is very important. We also put forward that the 3PDF should be adopted by the SE community to keep relevant evidence up to date and to avoid wasting effort with unnecessary updates.

One of the avenues of future work is to also contact authors of SE SLRs, which have not been updated yet, in order to obtain their assessment on the suitability of the 3PDF for use in SE.

# Appendix A

Seven excluded studies listed in Table 4.

# Appendix B

The survey questionnaire contains an introduction and three steps to provide feedback on each of the decision framework questions and decision drivers. The survey questions provided in each of the three steps, as presented to the respondents, are shown in Figure B.1.

### SLR Update Survey

* Required

Step 1 (out of 3): Assess an SLR's currency

Link to original framework:
https://www.dropbox.com/s/5z580abct9tg9bo/DecisionFramework.pdf

**1.a. Does the published SLR still address a current question (i.e., is the SLR current/actual in terms of its topic relevance for research and practice)?** *
- ○ This question is RELEVANT
- ○ This question is NOT relevant

Please justify your answer
Your answer

**1.b. Has the SLR had good access or use (i.e., has the SLR had an impact on research and/or practice)?** *
- ○ This question is RELEVANT
- ○ This question is NOT relevant

Please justify your answer
Your answer

**1.c. Has the SLR used valid methods and was it well conducted (i.e., has the SLR been carried out properly and using a sound methodology, e.g., with clear inclusion and quality assessment)?** *
- ○ This question is RELEVANT
- ○ This question is NOT relevant

Please justify your answer
Your answer

In the original Framework ALL questions in Step 1 MUST have a "YES" answer before moving on to Step 2. In other words, if at least one answer is "NO", then the SLR is judged as not needing to be updated.

**Would you agree that ALL questions in Step 1 MUST have a "YES" answer, before moving on to Step 2, when considering ONLY the questions in Step 1 that you judged as RELEVANT? If you do NOT AGREE, could you please provide a suggestion under 'Comments'?** *
- ○ Yes
- ○ No

Comments
Your answer

BACK    NEXT

### SLR Update Survey

* Required

Step 2 (out of 3): Identify relevant new methods, studies and other information

Link to original framework:
https://www.dropbox.com/s/5z580abct9tg9bo/DecisionFramework.pdf

**2.a. Are there any new relevant methods (examples of new methods can be new ways to search for evidence and to do the synthesis of evidence, inclusion of other types of study designs, new statistical techniques or improvements to existing techniques used for meta-analysis)?** *
- ○ This question is RELEVANT
- ○ This question is NOT relevant

Please justify your answer
Your answer

**2.b. Are there any new studies, or other information (e.g., grey literature papers, in case such types of studies are also being considered for inclusion)?** *
- ○ This question is RELEVANT
- ○ This question is NOT relevant

Please justify your answer
Your answer

In the original Framework AT LEAST ONE of the questions in Step 2 MUST have a "YES" answer before moving on to Step 3. In other words, if ALL answers are "NO", then the SLR is judged as up to date.

**Would you agree that AT LEAST ONE of the questions in Step 2 MUST have a "YES" answer before moving on to Step 3, when considering ONLY the questions in Step 2 that you judged as RELEVANT? If you do NOT AGREE, could you please provide a suggestion under 'Comments'?** *
- ○ Yes
- ○ No

Comments
Your answer

BACK    NEXT

### SLR Update Survey

* Required

Step 2 Assess the effect of updating the review

Link to original framework:
https://www.dropbox.com/s/5z580abct9tg9bo/DecisionFramework.pdf

**3.a. Will the adoption of new methods change the findings/conclusions or credibility?** *
- ○ This question is RELEVANT
- ○ This question is NOT relevant

Please justify your answer
Your answer

**3.b. Will the inclusion of new data/studies/information change findings/conclusions or credibility?** *
- ○ This question is RELEVANT
- ○ This question is NOT relevant

Please justify your answer
Your answer

In the original Framework AT LEAST ONE of the questions in Step 3 MUST have a "YES" answer, otherwise the SLR is judged as up to date.

**Would you agree that AT LEAST ONE of the questions in Step 3 MUST have a "YES" answer, when considering ONLY the questions in Step 3 that you judged as RELEVANT? If you do NOT AGREE, could you please provide a suggestion under 'Comments'?** *
- ○ Yes
- ○ No

Comments
Your answer

**If you were considering to update an SLR, would the questions presented in this survey, and the steps in which they are asked, be relevant to help you in your decision?** *
- ○ Yes
- ○ No
- ○ Not sure

Please comment
Your answer

Are there any other comments that you would like to make?
Your answer

BACK    SUBMIT

**Figure B.1: Survey steps and questions as presented to the respondents.**

# Appendix C

This Table provides evidence gathered from each of the SLR updates and used to answer step 3b in the 3PDF.

**Table C.1: Changes in SLR's conclusions? (Step 3B).**

| ID | Publication timespan | Changes in conclusions? | Evidence |
|---|---|---|---|
| S1 | 7 years | Yes | "Results: … Eleven [M2, M3, M4, M6, M7, M10, M12, M13, M14, M15, M16] out of the 17 pa-pers that we have examined propose new models…"<br>"Results: .. Among the 11 new models, we have identified a first group [M2, M3, M12] whose main concern is to address SRP in agile contexts …"<br>"Results: A second group of new models seems more concerned in proposing … For the remaining new models, …"<br>"Results: In the case of the new SRP models, we report three main categories of outputs …"<br>"Discussions: In contrast with the results reported in [14], we have found more proposals aimed to scale in presence of large sets of requirements, …"<br>"Discussions: "It is interesting to analyse the evolution of the SRP field by comparing the results from [14] and ours ... In our literature review, we have found 6 papers  ..."<br>"Discussion: Svahnberg et al. [14] found that most models focused on a limited set of input fac-tors, mainly hard constraints. Only a 57,1% of the reviewed models considered soft factors. In our study, 15 out of the 17 models (88,2%) do include soft factors …"<br>"Conclusion: … The main results (detailed in Section 4) show some progress with respect the previous proposals in the period 1997-2008 surveyed in [14], in particular: …" |
| S2 | 1 year | Yes | "Abstract: Researchers can use our results to find more detailed information about particular techniques, and to design comparable and reproducible experiments. ...."<br>"Introduction: We previously presented preliminary results from our SLR in a workshop paper (Ferrari et al., 2018a), with general classifications of primary studies. This paper greatly extends and updates the prior short paper, and makes four key contributions: • …"<br>Section 3.3. "Step 4–Post-publication update: Our set of selected studies was updated after our preliminary publication [54]."<br>"Section 6: This section presents more details about the measurements the studies used …"<br>"Conclusions: We updated and extended our prior paper (Ferrari et al., 2018a), analyzed the evolution of research on the topic, and summarized its underlying goals, techniques, metrics used, and results achieved …"<br>"Conclusions: Based on the research questions investigated in this work, we summarize several recommendations. …"<br>"Conclusions: We also found several studies that combined techniques to achieve substantial savings. For example, …" |
| S3 | 0 year | No | Original study: "Conclusion – ... a significant part of the companies perceived inclination towards for the innovation adoption process influenced by technological, organizational and environmental contexts ... it is expected that the results in this systematic literature review would help in establishing knowledge on how the companies should adopt and migrate to the cloud, how the cost-benefit relationship can be evaluated as well as providers can be selected. These findings can be a useful reference to develop guidelines for an effective use of cloud computing …" |
| | | | Update study: "Conclusion – ... a significant part of the companies perceived inclination towards the innovative adoption process influenced by technological, organizational and environmental contexts ... The results in this systematic literature review can help the development of guidelines to support newcomers companies to adopt and migrate to the cloud, how the cost-benefit relationship can be evaluated as well as the selection of providers …" |
| **S4** | 1 year | Yes | "Abstract: ... As a major finding of the review, the results strongly suggest the need for more practitioner-oriented and industry-driven empirical studies in the area of safety certification...."<br><br>"Introduction: .. we classify into a hierarchical taxonomy … The taxonomy includes 49 basic evidence types and is, to our knowledge, the most comprehensive classification of safety evidence built to date."<br><br>"Introduction: … Our study notably indicates that a large majority of the approaches surveyed have not been validated in realistic settings and thus provide little information about their practical utility …"<br><br>"Introduction: ... The work we present here extends an earlier conference paper [21]. The main extensions are: (1) the addition of a new data source, namely Google Scholar, (2) significant expansion of the description of the research method and the results; and (3) our domain analysis …"<br><br>"Related work: … There are a number of SLRs in the literature whose scope partially overlaps with ours, …  None of these specifically address the topic of evidence for safety."<br><br>"Results: With respect to the application domains and the safety standards referred to in the studies, we identified eight application domains and 16 safety standards."<br><br>"4.1 … RQ1: We created a taxonomy, shown in Fig. 2, for evidence types based on the various evidence examples, artefacts, tools and techniques found in the primary studies …"<br><br>"4.1 … RQ1: Our results indicate that the most frequent evidence types referred to in the literature are Hazards Cause Specification …"<br><br>"4.2 … RQ2: … we identified some technique for structuring safety evidence. We divide the techniques into three main categories …"<br><br>"4.3 … RQ3: … We identified techniques for evidence assessment … We classify these techniques into four categories …"<br><br>"4.4 … RQ4: … We identified several categories of general challenges and needs related to providing safety evidence information and to structuring and assessing the evidence …"<br><br>"Discussion: … The results from the review provide a general research-oriented view on evidence provision … We believe that this taxonomy is a useful reference to new researchers, helping them get better acquainted with the area"<br><br>"Discussion: … Alongside the taxonomy, our results concerning evidence structuring and assessment serve as useful input for future work on tool support …"<br><br>"Discussion: … We believe that more research on safety evidence at lower levels of abstraction (system type level and specific system level) is necessary …"<br><br>"Discussion: … The results regarding evidence structuring (RQ2) are useful for both research and practice to promote further work on managing large collections of evidence data …"<br><br>"Conclusions and Future Work: … One of the main outcomes of the SLR is a general taxonomy of<br>safety evidence type …"<br><br>"Conclusions and Future Work: … The SLR further examined and classified existing techniques for structuring evidence information into three categories …"<br><br>"Conclusions and Future Work: …  the paper presented a comparison of eight safety-critical domains …"<br><br>"Conclusions and Future Work: … The SLR provides useful insights for both researchers and practitioners…" |

| S5 | 2 years | No | "Abstract: The results of this review, like our previously conducted systematic review (Sulayman and Mendes, 2009), show that very few studies have specifically focused on SPI …" |
|---|---|---|---|
| | | | "Research Gaps: … Similar to our previous execution …, the number of studies that met our inclusion criteria is small …" |
| | | | "Research Gaps: … Studies S1, S4, S5, S6, S7 and S8 propose generic SPI models … However, none of the studies has suggested a specific model for Web SPI …" |
| | | | "Research Gaps: …The systematic review did not identify any SPI model or technique specifically customized for Web companies … However, this should not be understood as indicating that there is no need for specific Web SPI models and techniques. In fact, previous work on Web engineering may suggest the contrary." |
| | | | "Research Gaps: … We argue that in terms of SPI it may also be the case that research on result oriented, cheaper and less time consuming SPI strategies for small and medium Web companies need to be conducted …" |
| | | | "Conclusion: Due to our strict inclusion criteria, the number of relevant studies found was small but the overall search process was very comprehensive …" |
| | | | "Conclusion: Found studies suggested did not suggest any specific model or technique made to measure for the SPI of Web companies …" |
| | | | "Conclusion: The review as mentioned earlier helped us in identifying the possible research gaps and directions." |
| S6 | 4 years | Yes | "Introduction: … Like the previous review, the current review recognises the heterogeneity of the research in this relatively new field and aims to provide some organisation and structure to this area." |
| | | | "Results: 18 papers described online games both for entertainment (9) and learning (9), suggesting that these have become more popular since the previous review in 2012, when only 9 were found." |
| | | | "Results: 21 studies in the current review provided additional evidence to support previous research documenting the attentional …" |
| | | | "Results: … the current update to the systematic review of Connolly et al. found many more papers reporting empirical evidence of the positive outcomes of playing games (512) than the previous review (129). This illustrates the increased interest in the positive impacts of digital games during the five year period from 2009 to 2014 compared to the previous five year period." |
| | | | "Discussion: More high quality papers using the term "serious games" were identified in the current review (34) than in the previous review (8) suggesting that the term has entered the mainstream in recent years …" |
| | | | "Discussion: As in the previous review, similar numbers of papers about entertainment games and games for learning were found, but these studies differed in their aims, the platform used for delivery and the learning and behavioral outcomes they addressed." |
| | | | "Discussion: ... Some progress is being made in understanding how specific game features engage players and support learning." |
| | | | "Discussion: Progress has been made too in understanding features that make games more engaging ..." |
| S7 | 5 years | Yes | "Introduction: This study contributes to the cloud evaluation literature by: - providing an updated review and synthesis of the CSEM literature between 2006 and 2016 identifying research deficiencies of CSEMs for suitably positioning new research activities …" |
| | | | "Related Work: We have improved our preliminary SLR by (1) extending the publication period and defining an improved search string to provide an increased number of retrieved related papers, (2) applying a new data extraction dimension to provide more useful information regarding the CSEM components, and (3) examining the CSEMs based on MCDM making the review more focused. This focus will facilitate building even closer ties between the cloud evaluation context |
| | | | and MCDM context by identifying and understanding the associated MCDM method problems and deficiencies."<br>"Conclusion: Presenting new lines of inquiry that could inspire researchers to conduct more research within the area of cloud service evaluation is a main contribution of this SLR …"<br>"Conclusion: These research deficiencies in CSEMs emphasize the requirement to rethink the proposed methods, components, and processes in evaluating cloud services." |
| **S8** | 3 years | Yes | "Abstract: … we discovered 31 different CSFs for ERP system implementation projects."<br>"Motivation: … several case studies, surveys, and literature reviews have already been conducted by different researchers (e.g., [13–15]). Most of these literature<br>reviews cannot be reproduced, because descriptions of the review methods and procedures are lacking …"<br>"Critical Success …: The goal of the performed literature review is to gain an in-depth understanding of the different CSFs already identified by other researchers."<br>"Literature Review – Results: In most previous literature reviews, the CSFs were grouped more coarsely so that a lower number of CSFs was used … we used a larger number than earlier researchers had because we expected the resulting distribution to be more insightful."<br>"Literature Review – Results: Comparing these results with other literature reviews, …, it became obvious that the top five factors were similar while only the ranks differed. Due to our large literature base, our total numbers are much higher ..."<br>"Literature Review – Results: … This clearly is a research gap in the ERP CSF research area …"<br>"Conclusion: Our study reveals that there are several papers – case studies, surveys as well as literature reviews – that are focusing on CSFs."<br>"Conclusion: … we derived 31 different CSFs. We identified the following top three CSFs …"<br>"Conclusion: This ranking is similar to the ranking of other literature reviews (e.g., [2, 14]).<br>Compared to these other reviews, the number of papers included in our study exceeds<br>their numbers. …" |
| **S9** | 7 years | Yes | "Partial Results: This difference relates to an increase in the number of cross-company models to at least two … This was not observed previously, when the original SLR was carried out."<br>"Discussion: ... in addition to the within-company dataset sizes may have influenced the changes in the patterns that were previously documented in the original SLR on which this research is based upon …"<br>"Discussion: … Contrary to the results detailed in [13], our results provide some support to that supposition, thus suggesting that time spent in quality control …"<br>"Conclusion and Future Work ...We believe that this newly observed behavior might be related to the …" |
| **S10** | 4 years | Yes | "Abstract: We identified a strong indication of solutions based on Artificial Intelligence ...  We also identified that there is a gap in terms of agreement on suitable cost drivers ... "<br>"Abstract: This updated review of the state of the art resulted in 24 new relevant papers selected …"<br>"Introduction: … we applied the Forward Snowballing technique [8] to ﬁnd out relevant studies since the reference review of Usman et al. [2]."<br>"Introduction: As contribution of our review, 24 new relevant papers were selected. Some findings from the reference review remain actual, but other questions have been raised in our research ..." |
| | | | "Conclusion: Some considerations of the reference SLR are still valid and current. However, we identified new trends." |
| **S11** | 3 years | Yes | "Abstract: The main contribution of this study relates to the identification of other practices, including sharing of the "social conscience", …"<br>"Introduction: This study extended the studies on this subject, considering from 2012 to 2014 year and also including 11 new studies found in current literature…"<br>"Results: Collaborative Technology practice was the most mentioned in both SLR. So, it was selected as an important practice."<br>"Results: Another limitation cited by the Arshad et al. (2012) is the …<br>This finding was supported in the study case presented in Zahedi and Babar (2014), where an attempt …"<br>"Results: Another practice detected in the selected works is Knowledge Sharing, …<br>This practice is also mentioned in the KM literature as a way for knowledge conversion known as "socialization" …"<br>"Conclusion: The study identified a 35% increase in volume related to KM practices in GSD if compared to the primary study …"<br>"Conclusion: It was also observed that the practices "common understanding" and "knowledge sharing" are even more frequent in the work of the past 12 years…"<br>"Conclusion: … Among the contributions of this work are the identification of KMP Shared social context, and the remark about the need and the importance to use new approaches to improve informal communication …" |
| **S12** | 3 years | Yes | "Discussion: An interesting finding of the SR is that no predictors of maintainability or factors that impact upon maintainability were found that ..."<br>"Discussion: The comparison of the two SRs suggests that findings differ in terms of use of maintainability prediction techniques, importance of the predictors, and the project stage when the predictors are gathered."<br>"Discussion: Therefore, one can argue that the comparison given in Section IV would have<br>yielded different results if the previously conducted SR was extended to include recently published studies." |
| **S13** | 3 years | No | "Abstract: ... There is no significant advance in the last 5 years …"<br>"RQ1: … The Original Study reached similar results…"<br>"RQ2: … The only new de-motivator that appeared in this study was …"<br>"RQ3: … Only one external sign was added to the original list …"<br>"RQ4: … All aspects identified in the current study are described in the original study, but two: … are cited in only one paper each …"<br>"RQ5: … The MOCC model ... constructed based on the findings from the original study, has been cited twice."<br>"Discussion: The findings of our review confirm the results from the original study …"<br>"Discussion: … several motivators in V did not appear as motivators in the original study."<br>"Discussion: Regarding the motivating aspects of software engineering, the current study agrees with the original study, and adds ... as motivating aspects …"<br>"Discussion: (JCT)' is the most recurrent model in the original study, but it was found only in two papers in the current study…"<br>"Discussion: Our results suggest that the overall understanding of how software engineers are actually motivated does not seem to have significantly advanced in the last five years …"<br>"Discussion: In both the current and the original studies, survey research represents the majority of empirical primary studies."<br>"Limitations of this review: The problem of cultural bias faced in the Original Study is also present in this study …"<br>"Results: This study has identified new descriptors for software engineers' characteristics, and new motivators and demotivators for software engineers, besides |
| | | | those described in the Original Study." |
| **S14** | 0 year | Yes | "Abstract: In this paper, we extend a systematic literature review (SLR) to cover decisions with respect to model driven development (MDD) tooling. We extract tooling information from the selected SLR papers and classify these software tools supporting DSML development ..."<br><br>"Section 3: In this section, we report on the results of the SLR regarding DSML tooling support. ..."<br><br>"Section 3.2: In previous work ([12], [14]), for each DSML design found documented in the publications of our paper corpus, we identified and recorded the decision options for six decision points (see Section 2.2). ..."<br><br>"Section 3.2: From the above sources, we extracted tooling information for each of the eight categories (see Section 3.1) ..."<br><br>"Section 3.2: In Table 3, we differentiate between open- and closed-source license models per tooling category ..."<br><br>"Conclusion: In this paper, we presented an extension to a SLR study to include an additional decision point of tooling support for UML-based DSML development ..."<br><br>"Conclusion: These empirical findings complement recent, qualitative research on MDD tool adoption [9]." |
| **S15** | 1 year | No | "Abstract: The review identified 42 primary studies from 1992 to 2013. … The overall findings confirmed the positive impact of SPS on education"<br><br>"Introduction: To achieve the goal, we extended our previous Systematic Literature Review (SLR) [15, 14, 5] on this more specific topic …"<br><br>"Search Process: Our previous work [5] covered non-game simulation to a large extent, … Hence, there was a need to carry out a search using the identified important games."<br><br>"Related Work: In summary, our work differs from the previous related endeavors in a number of aspects. In particular, we set a different review scope from others with special focus on process simulation and explicitly include non-game simulators; we paid special attention on how to support SE educators to successfully adopt SPS in their curricula, which was rarely raised before."<br><br>"Conclusions: To be specific, it produces the following contributions to the SPS and SEE communities: … It presents a most recent update and exhaustive collection of the research on this growing topic; … larger number of relevant studies were identified than ever …; The review scope first explicitly combined the simulation games and non-game simulators …; The existing SEE simulators were described with their individual capabilities …; An initial version of the recommendation checklist was proposed …" |
| **S16** | 3 years | Yes | "Abstract: Our analysis reveals a field that is rapidly growing, both in volume and empirical focus, while becoming more mature."<br><br>"Abstract: We identify signs of field maturity from the increase in: (i) the number of journal articles, (ii) the empirical models within the last two years, and (iii) the number of ecosystems studied."<br><br>"Introduction: Our analysis identifies two main issues that inhibit the evolution of the field, …"<br><br>"Introduction: … our analysis indicates that research has evolved outside the borders of the existing definitions …"<br><br>"Analysis: We notice that the number of publications has been increasing since 2007."<br><br>"Analysis: We notice that the tendency of increased publications along the years is also followed by a tendency of increased journal papers…."<br><br>"Analysis: We note that there is an increase in the percentage of the groups 'report' and 'empirical model', while |
| | | | there is a decrease in 'tools or notation', 'qualitative or descriptive models', 'analytic models', and 'specific solution'."<br>"Analysis: We find that there are some new concepts that appear in the literature, …"<br>"Analysis: .. we identify some emerging concepts in the new literature …"<br>"Discussion: we note that the field has been increasing significantly in volume …"<br>"Summary: We document the evolution in the field by analyzing the literature as a time sequence …"<br>"Summary: our analysis shapes the view of the field having evolved outside the existing definitions of software ecosystems and, thus, propose the update of the definition of software ecosystems." |
| S17 | 2 years | Yes | "Abstract: The findings show that quantitative studies and the theoretical underpinnings of agile ISD are lacking."<br>"Abstract: In consequence, multiple research gaps that need further research attention are identified."<br>"Discussion: This review extends findings of previous reviews [14, 15, 17] by introducing new perspectives and by including contributions until early 2013."<br>"Discussion: Furthermore, this review exposed promising research areas by presenting a systematic map on the focus of the studies and the employed research methods."<br>"Discussion: To sum up, this review serves as foundation for future studies by opening up several research possibilities …" |
| S18 | 8 years | Yes | "Abstract: Key results include that in contrast to 1999–2009,"<br>"Research Motivation: … although there have been reviews covering different aspects of agile GSD in the past, we have identified the research gap that there is no recent study focusing …"<br>"Research Motivation: … This study is furthermore unique in the way that it effectively covers the usage of agile practices in GSD from agile origins (1999) to recent time (2016)."<br>"Research Motivation: … we see particular value in our study by continuing the work of Jalali and Wohlin (S62; S63) to perform a complete and updated analysis in contrast to covering the whole time frame on our own"<br>"Results: Compared to years 1999–2009 (S63), there is an increase in total paper count per year …"<br>"Results: we can see a maturing research field with a shift from experience reports, …"<br>"Results: a decrease in the usage of purely qualitative methods in favor of more mixed and quantitative approaches."<br>"Results: there is a rise in literature reviews …"<br>"Results: although overall the most frequently seen contribution (51%), have decreased drastically<br>from 70% in 1999–2009 to 20% in 2010–2016 in favor of more rigorous contributions such as case study analysis …"<br>"Results: For 2010–2016, although most studies still lack some details of the empirical context, …"<br>"Results: In our newer studied period of 2010–2016, the customer has been even less frequently reported and remained unclear in 65% of relationships …"<br>"Results: From 1999 to 2016, the "standup meeting" has been most frequently applied (70.5 cases), …"<br>"Discussion: We have found a great variety of distribution scenarios in years 2010–2016 and were able to provide deeper insight compared to (S63), …"<br>"Conclusion: The research contribution of this study lies in reporting the state of the art of agile practices in GSD for 1999–2016, where we almost exclusively found success stories."<br>"Conclusion: We also saw several modifications and extensions of agile practice in use to deal with the challenges of GSD." |

| **S19** | 3 years | No | "Abstract: This paper presents recommendations .... Recommendations are based on a previous systematic review, which was updated and expanded with 13 new empirical studies and more than 60 new empirical results."<br>"Abstract: ... it was found that interviews were one ... The new evidence supports the same conclusion."<br>""Main findings: The sensitivity analysis had been repeated in our updating SR ... The final result is the same. ..."<br>"Main findings: ... it was not possible to corroborate the similarity of [3] and [12] due to the fact that ..."<br>""Discussion: The newly obtained evidence points out in the same direction than the previously existing one."<br>""Discussion: ... the present study gathers more evidence for the effectiveness of interviews and laddering."<br>"Discussion: The inferiority of sorting techniques is also corroborated by the present study ..."<br>"Discussion: The present study also confirms that the efficacy of interviews, laddering, ..."<br>"Discussion: ... we did not find new evidence about a new technique (scaling) not considered before."<br>"Discussion: ... it is noticeable that only 3 of the 9 papers reviewed .... have contributed to the set of aggregations ..."<br>"Discussion: Quite the contrary, the same fact was already observed in the previous review ..."<br>"Conclusions: We already established in [16][17] that interviews were able to extract more information ..."<br>"Conclusions: ... In the present work, we could not find any contradictory evidence ... the results corroborate the superiority of interviews ..." |
|---|---|---|---|
| **S20** | 6 years | Yes | "Abstract: Compared the original study, the larger group of empirical studies of the UML through our analysis concentrates on comprehension"<br>"Introduction: we mainly focus on the change and the development trend in the research of UML in recent years by means of extension into an SLR ..."<br>"Discussion: had a tendency to continue for future studies (this validate the prediction of the original research). ..."<br>"Discussion: The number of empirical studies focuses on UML metrics has presented a significant decreasing ..."<br>"Discussion: Not surprisingly, the type of diagram that is most frequently studied show a clear sequence: class diagrams …"<br>"Discussion: This result confirms one of the points of the several related work and original study …"<br>"Discussion: Although the original study [3] has failed to show a direct statement, we could still derive the general trend …"<br>"Conclusion: we had presented an extension of a systematic literature review for assessing the further development trend of the notations of the UML and the effectiveness in practical application ..."<br>"Conclusion: We also should note more studies are also necessary conducted with professional software engineers or practitioners to raise the generalizability of the results." |